\documentclass[pra,reprint,twocolumn,longbibliography,superscriptaddress]{revtex4-2}

\usepackage{amsmath,amssymb,bm,mathtools}
\usepackage{physics}
\usepackage{graphicx}
\usepackage{microtype}
\usepackage[section]{placeins}
\usepackage[colorlinks=true,linkcolor=blue,citecolor=blue,urlcolor=blue]{hyperref}

\newcommand{\halfline}{\mathbb{Z}_{\ge 0}}
\newcommand{\Hil}{\mathcal{H}}
\renewcommand{\dd}{\mathrm{d}}
\newcommand{\ii}{\mathrm{i}}
\newcommand{\1}{\mathbf{1}}
\newcommand{\E}{\mathbb{E}}

\newcommand{\opn}{\mathrm{op}}

\renewcommand{\Tr}{\mathrm{Tr}}

\begin{document}

\title{Maximal-velocity deficit under a finite-support constraint in a hard-wall half-line continuous-time quantum walk}

\author{Kangqiao Liu}
\email{kqliu@xhu.edu.cn}
\affiliation{School of Science, Xihua University, Chengdu 610039, China}
\affiliation{Key Laboratory of High Performance Scientific Computation, Xihua University, Chengdu 610039, China}
\author{Deyou Chen}
\email{deyouchen@hotmail.com}
\affiliation{School of Science, Xihua University, Chengdu 610039, China}
\affiliation{Key Laboratory of High Performance Scientific Computation, Xihua University, Chengdu 610039, China}

\date{\today}

\begin{abstract}
Continuous-time quantum walks on a lattice spread ballistically and converge to a limiting distribution for the rescaled position.
On the hard-wall half line the boundary reflects the walker but does not change the bulk dispersion, so the ballistic front remains set by the maximal group velocity.
We ask how close one can get to this front when the initial state is constrained to occupy only the first $M$ sites.
We show that optimizing the moment-generating function of the limiting-velocity distribution over this finite-support class reduces to the principal eigenvalue of an explicit $M\times M$ Hermitian matrix, which yields the optimal state by direct diagonalization.
For large $M$, a near-front scaling limit produces a continuum description that controls the entire peak region.
In particular, the maximal mean drift approaches the front with an inverse-square deficit in $M$ and a constant prefactor $\pi^2/4$.
Numerical results verify both the finite-$M$ spectral formulation and the predicted scaling.
\end{abstract}

\maketitle

\section{Introduction}
\label{sec:intro}

Continuous-time quantum walks (CTQWs) provide a minimal setting where coherent interference drives ballistic rather than diffusive transport on discrete structures \cite{FarhiGutmann1998DecisionTrees,Kempe2003IntroOverview,MulkenBlumen2011PhysRepCTQW,VenegasAndraca2012ComprehensiveReview}.
Recent work has connected CTQWs to algorithmic tasks such as spatial search and state transfer and has realized programmable continuous-time dynamics in photonic and atomic platforms \cite{ApersChakrabortyNovoRoland2021QuadraticSpeedup,Gong2021Science62Qubit,Young2022ScienceTweezerQW,Qiang2021SiliconPhotonicQW}.
These advances motivate a quantitative understanding of how preparation resources bound the transport that can be achieved.

For the translation-invariant nearest-neighbor walk on the line with hopping amplitude $J>0$, the dispersion is $E(k)=-2J\cos k$ and group velocities fill $[-2J,2J]$.
A stationary-phase argument links ballistic rays $X_t\approx vt$ to the group-velocity matching condition $v=E'(k)$, where $t$ denotes the evolution time. For initial states with a regular momentum profile on the full line, the distribution of $X_t/t$ converges to a limiting velocity distribution induced by the group-velocity map \cite{Konno2005,Gottlieb2005,CedzichJoyeWernerWerner2025Tail}.
The maximal front speed is $2J$, attained at the dispersion peak $k_0=\pi/2$.

We study the hard-wall half line $\halfline=\{0,1,2,\dots\}$, where a boundary at the origin enforces reflection and replaces plane waves by standing-wave modes \cite{Machida2024HalfLineCTQW}.
A common theme in local quantum dynamics is that propagation has a finite speed, a fact formalized for many-body systems by Lieb-Robinson type bounds \cite{LiebRobinson1972FiniteVelocity,Tran2021LRLightConePowerLaw,WangHazzard2020TighteningLR,ChenLucasYin2023SpeedLocalityReview}.
In our single-particle walk the same light-cone picture appears in its simplest form. The bulk dispersion fixes the maximal ballistic velocity $2J$, while the hard wall does not change this value.
The question is therefore how finite-support initial states restrict how closely one can approach $2J$ in velocity space.
Approaching $2J$ requires concentrating spectral weight in an increasingly narrow window around $k_0=\pi/2$, and finite control near the boundary makes this concentration a nontrivial optimization problem.

We implement the preparation constraint by restricting the initial state to the first $M$ sites.
This finite-support constraint captures the practical limit of local addressing and turns the optimization into a finite-dimensional problem.
We focus on the maximal mean drift velocity under this constraint.
Let $V$ be the asymptotic velocity random variable defined by the weak limit $X_t/t \Rightarrow V$ as $t\to\infty$.
We quantify the best achievable ballistic transport at support size $M$ by the optimal mean drift $v_{\max}(M)$,
and ask how the deficit $2J-v_{\max}(M)$ decays as $M$ increases. We will use an exponential-tilting moment generating function (MGF) $G_\psi(s)=\E_\psi[e^{sV}]$ as a tool \cite{Touchette2009LargeDeviationsReview,ChetriteTouchette2015Conditioned}.

Ultracold atoms in optical lattices provide a direct implementation of tight-binding Hamiltonians and have been used to observe quantum walk in situ \cite{Preiss2015StronglyCorrelatedQW,GrossBakr2021QGM}. Site-resolved detection and programmable optical potentials in quantum-gas microscopes allow steep barriers and spatially confined preparations. These controls approximate the hard-wall half-line geometry and the ideal finite-support constraint \cite{GrossBakr2021QGM,DiCarli2024CommensurateIncommensurate}.

Our results have two parts.
For each fixed $M$, evaluating the optimized MGF over the finite-support class reduces to the principal eigenvalue and eigenvector of an explicit $M\times M$ matrix.
This yields the optimal state and the optimal value by direct diagonalization.
We then study the near-front regime where the limiting velocity approaches $2J$.
In this regime the constraint to $M$ sites resolves only a momentum window of width $\Delta k\sim 1/M$ around $k_0=\pi/2$.
After rescaling within this window, the discrete optimization converges to a continuum eigenproblem governed by a compact self-adjoint operator $T_u$ built from a sharp Fourier cutoff and a Gaussian real-space window.
Its principal eigenvalue defines a scaling function $\lambda(u)$ that controls the full near-peak statistics.
Expanding $\lambda(u)$ near $u=0$ yields the constant $\pi^2/4$ that controls the asymptotic defect $2J-v_{\max}(M)$.

In Sec.~\ref{sec:model} we define the hard-wall CTQW and introduce the half-line sine diagonalization.
Section~\ref{sec:ballistic} summarizes the ballistic scaling limit and the induced limiting velocity distribution on the half line.
Section~\ref{sec:finiteM} gives the finite-$M$ spectral characterization of the optimal biased statistics.
Section~\ref{sec:continuum} derives the near-front scaling limit and the associated continuum operator.
Section~\ref{sec:constant} extracts the constant controlling the optimal approach to the ballistic front.
Section~\ref{sec:numerics} presents numerical confirmation.
We present the experimental relevance of our setup and results in Sec.~\ref{sec:exp_relevance}. Implications and extensions are discussed in Sec.~\ref{sec:discussion}.

\section{Hard-wall half-line CTQW}
\label{sec:model}
We work on the Hilbert space $\Hil=\ell^2(\halfline)$ with the orthonormal basis $\{\ket{n}\}_{n\ge 0}$.
The hard-wall nearest-neighbor Hamiltonian is \cite{FarhiGutmann1998DecisionTrees,MulkenBlumen2011PhysRepCTQW,Machida2024HalfLineCTQW}
\begin{equation}
H=-J\sum_{n\ge 0}\left(\ket{n+1}\bra{n}+\ket{n}\bra{n+1}\right),
\label{eq:H}
\end{equation}
where the hopping amplitude satisfies $J>0$.
The state evolves as $\ket{\psi(t)}=e^{-\ii tH}\ket{\psi}$.
We denote the position operator by $X=\sum_{n\ge 0}n\ket{n}\bra{n}$ and the position random variable by
\begin{equation}
\mathbb{P}_\psi(X_t=n)=\abs{\ip{n}{\psi(t)}}^2,
\end{equation}
where $X_t$ denotes the outcome of measuring $X$ in the state $\ket{\psi(t)}$. 

A hard wall corresponds to Dirichlet boundary conditions at site $-1$ and leads naturally to a sine spectrum.
Define the unitary sine transform $\mathcal{F}_s:\ell^2(\halfline)\to L^2((0,\pi),\dd k)$ by 
\begin{equation}
(\mathcal{F}_s\psi)(k)=\sqrt{\frac{2}{\pi}}\sum_{n\ge 0}\psi_n\sin[(n+1)k] =: c_{\psi}(k),
\label{eq:sine}
\end{equation}
where we write $\psi_n=\ip{n}{\psi}$ and the subscript $s$ indicates `sine'.
This sine representation is the half-line analogue of momentum space, adapted to the Dirichlet boundary at the wall.
Applying the unitary sine transform $\mathcal{F}_s$, the Hamiltonian becomes diagonal in the $k$ representation, i.e., $\mathcal{F}_s H\mathcal{F}_s^{-1}$ acts as multiplication by the dispersion relation
\begin{equation}
E(k)=-2J\cos k.
\end{equation}
The associated group velocity is
\begin{equation}
v(k)=\frac{\dd E}{\dd k}=2J\sin k\in[0,2J].
\label{eq:vg}
\end{equation}
The maximal velocity $2J$ is attained at $k_0=\pi/2$ where the group velocity has a quadratic peak:
\begin{equation}
2J-v\left(\frac{\pi}{2}+q\right)=Jq^2+O(q^4).
\label{eq:quadpeak}
\end{equation}
This expansion holds as $q\to 0$.
This peak curvature is the origin of the $M^{-2}$ defect scaling derived below.

The bulk dispersion is unchanged, so ballistic scaling is still controlled by how the initial state distributes spectral weight over $k\in(0,\pi)$ under the velocity map $v(k)=2J\sin k$.
The hard wall enters only through the standing-wave form factors in \eqref{eq:sine}, which modify the near-boundary interference pattern but do not change the velocity interval.
We recall the corresponding limiting-velocity distribution in Sec.~\ref{sec:ballistic} following standard CTQW weak-limit analyses \cite{Konno2005,Gottlieb2005,MulkenBlumen2011PhysRepCTQW,CedzichJoyeWernerWerner2025Tail}.

\section{Ballistic scaling and the limiting velocity distribution}
\label{sec:ballistic}
\subsection{Stationary-phase analysis}

Let $w_\psi(k)=\abs{c_\psi(k)}^2$.
For any normalized $\psi$ with a finite support, $w_\psi(k)\dd k$ is a probability measure on $(0,\pi)$.
The ballistic weak limit on the half line can be stated in the following change-of-variables form \cite{Machida2024HalfLineCTQW}. 
For any bounded continuous $f:\mathbb{R}\to\mathbb{R}$,
\begin{equation}
\lim_{t\to\infty}\E_\psi\left[f\left(\frac{X_t}{t}\right)\right]
=\int_0^\pi w_\psi(k)f\left(v(k)\right)\dd k.
\label{eq:ballistic_limit}
\end{equation}
Closely related weak-limit theorems and asymptotic-velocity results on the full line can be found in Refs.~\cite{Konno2005,Gottlieb2005,CedzichJoyeWernerWerner2025Tail,richardII,MandalSarkarChakrabortyAdhikari2022PRA}. Related rigorous spectral and dispersive approaches for one-dimensional discrete-time quantum walks include, for example, eigenfunction-expansion methods and related unitary spectral representations \cite{Tate2022AFA} as well as dispersive estimates and sharp velocity bounds in inhomogeneous or periodically driven settings \cite{MaedaSasakiSegawaSuzukiSuzuki2022JMSJ,AbdulRahmanStolz2023CMP}.
Equivalently, there exists a random variable $V\in[0,2J]$ such that $X_t/t\Rightarrow V$ and the law of $V$ is induced from the quasi-momentum measure $w_\psi(k)\dd k$ by the mapping $k\mapsto 2J\sin k$.

Equation~\eqref{eq:ballistic_limit} shows that the long-time transport statistics are entirely controlled by how much spectral weight the initial state places at momenta with given group velocity.
In particular, any attempt to approach the maximal velocity $2J$ requires concentrating $w_\psi(k)$ near $k_0=\pi/2$.
The hard wall affects this concentration in two ways. First, the eigenmodes are standing waves, so the spectral weight is built from a sine polynomial rather than a full Fourier transform. Second, the finite-support constraint restricts the allowed near-peak profiles. 

The following stationary-phase sketch explains how the hard-wall sine spectrum produces the ballistic ray structure behind Eq.~\eqref{eq:ballistic_limit}.
The velocity interval is $[0,2J]$ and, for a given outward velocity $v\in(0,2J)$, there are two quasi-momenta that contribute to the same ray because $v(k)=2J\sin k$ is symmetric about $k_0=\pi/2$.
From the sine diagonalization,
\begin{align}
\psi_n(t)=\sqrt{\frac{2}{\pi}}\int_{0}^{\pi} e^{\ii 2Jt\cos k}c_\psi(k)\sin[(n+1)k]\dd k.
\label{eq:amp_sp}
\end{align}
For large $t$ we consider ballistic rays $n=\lfloor vt\rfloor$ with $v\ge 0$.
Writing
\begin{equation}
\sin[(n+1)k]=\frac{1}{2\ii}\left(e^{\ii(n+1)k}-e^{-\ii(n+1)k}\right),
\end{equation}
we isolate the phases
\begin{equation}
\phi_\pm(k)=2J\cos k \pm vk.
\end{equation}
Stationary points satisfy $\phi_\pm'(k)=0$, hence $-2J\sin k\pm v=0$ \cite{vanDerCorput1935StationaryPhaseI,Erdelyi1955StationaryPhase}.
This is the usual velocity-matching condition $v=v(k)$, which identifies the quasi-momenta whose contributions add coherently along the ray.
When no stationary point lies in the integration domain, the integral is dominated by oscillatory cancellation and is smaller than the stationary contribution, as quantified by standard nonstationary-phase bounds \cite{vanDerCorput1935StationaryPhaseI,Erdelyi1955StationaryPhase}.
For $v\ge 0$ only the $+$ branch has stationary points in $(0,\pi)$.
For $v\in(0,2J)$ there are two stationary points,
\begin{align}
k_1(v)=\arcsin\left(\frac{v}{2J}\right)\in\left(0,\frac{\pi}{2}\right),\ k_2(v)=\pi-k_1(v).
\end{align}
At $v=0$ the stationary points reach the boundary, and at $v=2J$ they merge at $k_0=\pi/2$ and become degenerate.
For the weak limit with bounded continuous test functions, the interior stationary phase yields the limiting density on $(0,2J)$ and the endpoint analysis fixes the support $[0,2J]$.

For $v$ bounded away from $0$ and $2J$, assume that $c_\psi$ is $C^2$ in neighborhoods of $k_1(v)$ and $k_2(v)$ and that $c_\psi$ and its first derivative are bounded on $[0,\pi]$.
This mild regularity is automatic for the finite-support states considered below since then $c_\psi$ is a finite trigonometric sum.
A standard one-dimensional stationary-phase estimate then gives \cite{vanDerCorput1935StationaryPhaseI,Erdelyi1955StationaryPhase}
\begin{align}
\psi_{\lfloor vt\rfloor}(t)=\frac{1}{\sqrt{t}}
\sum_{j=1}^{2} a_j(v)c_\psi(k_j(v))e^{\ii t \phi_+(k_j(v))+\ii\sigma_j}+o(t^{-\frac{1}{2}})
\label{eq:sp_amp_main}
\end{align}
with the standard stationary-phase prefactors $a_j(v)$, which are explicitly determined by $\phi_+''(k_j(v))$ and are nonzero for $|v|<2J$, and the phase factors $\sigma_j=\mathrm{sgn}(\phi_+''(k_j(v)))\pi/4$.
Taking the modulus square produces branch terms of order $t^{-1}$ and oscillatory cross terms with phase $t(\phi_+(k_1(v))-\phi_+(k_2(v)))$.

To identify the weak limit, we use the Born rule and test against $f\in C_c^\infty(\mathbb{R})$.
\begin{equation}
\E_\psi\left[f\left(\frac{X_t}{t}\right)\right]=\sum_{n\ge 0} f\left(\frac{n}{t}\right)\abs{\psi_n(t)}^2.
\label{eq:test_main}
\end{equation}
To pass from the ray-wise stationary-phase expansion \eqref{eq:sp_amp_main}, valid for each fixed $v\in(0,2J)$, to the weak limit \eqref{eq:ballistic_limit}, we insert \eqref{eq:sp_amp_main} into the test average \eqref{eq:test_main}.
The cross term is an oscillatory contribution with a $t$-dependent phase and averages out as $t\to\infty$ by standard oscillatory-cancellation estimates, while the branch contributions yield the change-of-variables law induced by the mapping $k\mapsto v(k)=2J\sin k$.
The cancellation step can be justified using standard nonstationary-phase bounds \cite{vanDerCorput1935StationaryPhaseI,Erdelyi1955StationaryPhase}.
In the remainder of this paper we take \eqref{eq:ballistic_limit} as the ballistic input and quantify, under the finite-support constraint, how close the optimized statistics can approach the velocity peak.

\subsection{Boundary-localized state and exact mean velocity}
\label{subsec:example_boundary_state}

A simple example already illustrates how the hard wall shapes transport statistics and connects directly to the weak-gradient saturation reported recently in Maxwell-demon-assisted transport \cite{LiuNakagawaUeda2023MaxwellDemonTransport}.
Consider the boundary-localized initial state $\ket{\psi}=\ket{0}$ at the site adjacent to the wall.
We write $\E_\psi[V]$ for the mean limiting velocity.
In this example we suppress the state subscript and write $c(k)$ and $w(k)$ for $c_\psi(k)$ and $w_\psi(k)$.
Since the sine eigenmodes are $\varphi_n(k)=\sqrt{2/\pi}\sin[(n+1)k]$, we have
\begin{equation}
\begin{aligned}
c(k)&=(\mathcal{F}_s\ket{0})(k)=\sqrt{\frac{2}{\pi}}\sin k,\\
w(k)&=\abs{c(k)}^2=\frac{2}{\pi}\sin^2 k.
\end{aligned}.
\label{eq:w_boundary}
\end{equation}
The ballistic weak limit \eqref{eq:ballistic_limit} then yields the mean limiting velocity,
\begin{align}
\E_\psi[V]=\int_0^\pi w(k)v(k)\dd k=\frac{4J}{\pi}\int_0^\pi \sin^3 k\dd k
=\frac{16J}{3\pi}.
\label{eq:vbar_boundary}
\end{align}
For $J=1$ this gives $\E_\psi[V]=16/(3\pi)\approx 1.69765$, quantitatively explaining the saturated value $\simeq 1.7$ observed in the weak-gradient regime of Ref.~\cite{LiuNakagawaUeda2023MaxwellDemonTransport}.

In the remainder of this work we go beyond this fixed initial state and ask, under a finite-support constraint, which initial states optimize the limiting-velocity MGF \eqref{eq:MGF} and thereby, in particular, the maximal mean drift.

\section{Finite-support resource constraint and exact spectral optimization}
\label{sec:finiteM}
\subsection{Velocity MGF under finite support}

Fix an integer $M\ge 1$ and impose the finite-support constraint
\begin{equation}
\mathcal{S}_M=\left\{\psi\in\ell^2(\halfline):\psi_n=0,\ \forall n\ge M,\ \sum_{n=0}^{M-1}\abs{\psi_n}^2=1\right\}.
\label{eq:SM}
\end{equation}
Physically, $M$ is the size of the boundary region in which the initial state can be prepared or controlled.
Mathematically, $\mathcal{S}_M$ identifies with the unit sphere of $\mathbb{C}^M$, making optimization problems finite dimensional.
For $\psi\in\mathcal{S}_M$ define the limiting-velocity MGF as
\begin{align}
 G_\psi(s)&=\E_\psi[e^{sV}]=\int_0^\pi w_\psi(k)e^{sv(k)}\dd k\nonumber\\
 & =\int_0^\pi w_\psi(k)e^{2Js\sin k}\dd k.
 \label{eq:MGF}
\end{align}
Because $V\in[0,2J]$, $G_\psi(s)$ is finite for all real $s$ and satisfies $1\le G_\psi(s)\le e^{2Js}$ for $s\ge 0$.
The parameter $s$ exponentially tilts the velocity distribution, so optimizing $G_\psi(s)$ probes the best achievable weight in the high-velocity tail.

Mean drift corresponds to the derivative at $s=0$, namely $\E_\psi[V]=\partial_s G_\psi(s)\vert_{s=0}$.
Equation~\eqref{eq:sine} implies $c_\psi(k)=\sqrt{2/\pi}\sum_{n=0}^{M-1}\psi_n\sin[(n+1)k]$.
Inserting into Eq.~\eqref{eq:MGF} and expanding $\abs{c_\psi(k)}^2$ gives a quadratic form in the coefficient vector $\psi=(\psi_0,\dots,\psi_{M-1})\in\mathbb{C}^M$:
\begin{equation}
G_\psi(s)=\expval{B^{(M)}(s)}{\psi},
\label{eq:MGF_RQ}
\end{equation}
where $B^{(M)}(s)$ is the explicit $M\times M$ Hermitian matrix with entries given by
\begin{equation}
B^{(M)}_{mn}(s)=\frac{2}{\pi}\int_0^\pi \sin[(m+1)k]\sin[(n+1)k]
e^{2Js\sin k}\dd k,
\label{eq:Bms}
\end{equation}
where $0\le m,n\le M-1$. Therefore, the optimal MGF is the top eigenvalue,
\begin{equation}
G^{(M)}_{\max}(s)=\max_{\psi\in\mathcal{S}_M}G_\psi(s)=\lambda_{\max}\left(B^{(M)}(s)\right).
\label{eq:optMGF}
\end{equation}
The optimal initial states are precisely the normalized principal eigenvectors of $B^{(M)}(s)$.
This spectral reduction is exact and does not require any time evolution.
This observation will be used in the next subsection to reduce the mean-drift optimization to the top eigenvalue of a second explicit matrix.

\begin{figure}[t]
 \centering
 \includegraphics[width=\linewidth]{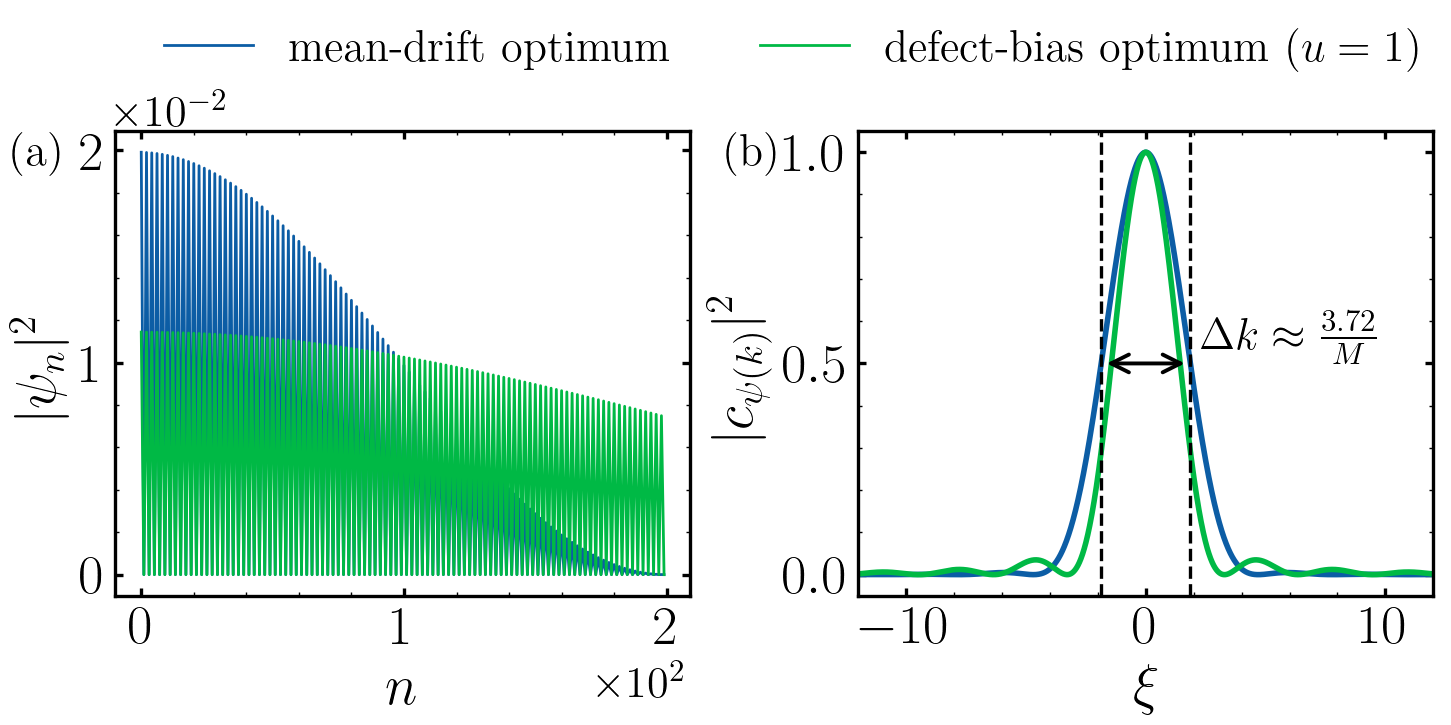}
 \caption{Optimal finite-$M$ states in (a) real space and (b) near the maximal-velocity peak.
We set $J=1$ and $M=200$.
The mean-drift optimum is the principal eigenvector of $A^{(M)}$ in Eq.~\eqref{eq:Adef}.
The defect-bias optimum is the principal eigenvector of $A_u^{(M)}$ in Eq.~\eqref{eq:AuM} at $u=1$.
(a) shows the real-space probability $\abs{\psi_n}^2$.
(b) shows the near-peak spectral weight $\abs{c_\psi(k)}^2$ as a function of the scaled deviation $\xi=M(k-\pi/2)$.
The indicated bandwidth is the full width at half maximum of the near-peak spectral profile, giving $\Delta k\approx 3.72/M$ at $M=200$.
Each curve is normalized to unit maximum to compare shapes.}
 \label{fig:optbias}
\end{figure}

\subsection{Mean drift and spectral optimization}
The maximal mean drift speed under the resource constraint is
\begin{equation}
v_{\max}(M)=\max_{\psi\in\mathcal{S}_M}\E_\psi[V].
\label{eq:vmax_def}
\end{equation}
Differentiating Eq.~\eqref{eq:MGF} at $s=0$ shows that the mean drift is given by
\begin{align}
 \E_\psi[V]=\expval{A^{(M)}}{\psi}
\end{align}
with
\begin{align}
 A^{(M)}_{mn}=\frac{2}{\pi}\int_0^\pi \sin[(m+1)k] \sin[(n+1)k] 2J\sin k \dd k.
 \label{eq:Adef}
\end{align}
Hence
\begin{equation}
v_{\max}(M)=\lambda_{\max}\left(A^{(M)}\right),
\label{eq:vmax_eig}
\end{equation}
and the mean-drift optimal initial state is the principal eigenvector of $A^{(M)}$.

The matrix $A^{(M)}$ has an explicit Toeplitz-Hankel closed form, meaning a sum of a bulk Toeplitz part that depends only on $m-n$ and a boundary-reflection Hankel part that depends only on $m+n$ \cite{Widom1976BlockToeplitzII,ClaeysGlesnerMinakovYang2021OrthEnsembles,StrangMacNamara2014ToeplitzHankel}.
Writing $a=m+1$ and $b=n+1$ and using the product-to-sum identity, we find
\begin{equation}
A^{(M)}_{mn}=\frac{2J}{\pi}\left[\mathcal{S}(\abs{m-n})-\mathcal{S}(m+n+2)\right].
\label{eq:Aclosed}
\end{equation}
Here the auxiliary integral $\mathcal{S}(c)$ is defined by 
\begin{equation}
\mathcal{S}(c):=\int_0^\pi \sin k\cos(ck)\dd k.
\label{eq:Jc_def}
\end{equation}
For integer $c\ge 0$ this evaluates to
\begin{equation}
\mathcal{S}(c)=\begin{cases}
0, & c\in 2\mathbb{Z}+1,\\[6pt]
\displaystyle \frac{1}{c+1}-\frac{1}{c-1}, & c\in 2\mathbb{Z}.
\end{cases}.
\label{eq:Jc_closed}
\end{equation}
This representation makes the hard-wall contribution transparent and allows the optimal initial state to be obtained directly by diagonalizing $A^{(M)}$ and taking its principal eigenvector.

The optimization in Eq.~\eqref{eq:vmax_eig} is over coherent amplitudes, not only over site occupations.
A simple site-diagonal reference makes this distinction explicit.
If the initial preparation is restricted to a site-diagonal mixture $\rho=\sum_{n=0}^{M-1}q_n\ket{n}\bra{n}$, then the mean drift is $\Tr(\rho A^{(M)})=\sum_n q_n A_{nn}^{(M)}$.  The site-diagonal optimum is therefore the largest diagonal element of $A^{(M)}$.
Using Eq.~\eqref{eq:Aclosed},
\begin{equation}
A_{nn}^{(M)}=\frac{4J}{\pi}\left[1+\frac{1}{(2n+1)(2n+3)}\right],
\label{eq:incoherent_diagonal}
\end{equation}
which is maximal at $n=0$.  We denote the resulting site-diagonal incoherent reference value by
\begin{equation}
v_{\rm inc}^{\max}=\frac{16J}{3\pi}.
\label{eq:incoherent_reference}
\end{equation}
This value is independent of $M$ and coincides with the coherent optimum $v_{\max}(1)$, where no coherence between distinct sites is available.  Thus, within the site-diagonal class, enlarging the allowed support does not improve the drift.  The inverse-square approach to the front uses coherent interference among the allowed site amplitudes, as reflected in the parity-selected structure of the mean-drift optimal finite-$M$ state discussed below.

\section{Near-maximal-velocity scaling and continuum operator}
\label{sec:continuum}
The previous section reduces the finite-support optimization at fixed $M$ to an explicit $M\times M$ eigenproblem.
To characterize how the constraint limits access to velocities close to $2J$, we zoom into an $O(1/M)$ quasi-momentum window around the peak $k_0=\pi/2$ and derive a continuum-operator description of the large-$M$ regime.

\subsection{Defect variable and joint scaling}

To zoom into the near-maximal-velocity regime, we define the rescaled velocity defect
\begin{equation}
\Delta_M=\frac{M^2}{J}(2J-V)\ge 0.
\label{eq:defect}
\end{equation}
The $M^2$-scaling is dictated by the quadratic peak \eqref{eq:quadpeak} and the finite-support spectral resolution $\Delta k\sim 1/M$:
if $k=\pi/2+\xi/M$ with $\xi=O(1)$, then $2J-V\approx J\xi^2/M^2$, so $\Delta_M$ remains $O(1)$.
With this rescaling, the near-front defect remains $O(1)$ as $M$ grows, so it is natural to probe it through its Laplace transform at fixed $u\ge 0$.

We optimize the Laplace transform of $\Delta_M$,
\begin{equation}
\mathcal{G}_M(u)=\max_{\psi\in\mathcal{S}_M}\E_\psi[e^{-u\Delta_M}],
\label{eq:GM}
\end{equation}
where $u\ge 0$.
This quantity is equivalent to the optimal MGF under the joint scaling $s=(u/J)M^2$:
\begin{equation}
\mathcal{G}_M(u)=e^{-2Js}G_{\max}^{(M)}(s).
\label{eq:link_MGF_Laplace}
\end{equation}
We have $\mathcal{G}_M(u)\in(0,1]$, since $G_\psi(s)\le e^{2Js}$ for all $\psi$, and the prefactor $e^{-2Js}$ removes this state-independent growth.

Using the same Rayleigh-Ritz reduction as in Sec.~\ref{sec:finiteM}, inserting \eqref{eq:ballistic_limit} and the sine expansion gives
\begin{equation}
\mathcal{G}_M(u)=\lambda_{\max}\left(A_u^{(M)}\right),
\label{eq:GM_eig}
\end{equation}
where the $M\times M$ matrix $A_u^{(M)}$ has elements
\begin{align}
  (A_u^{(M)})_{mn}=\frac{2}{\pi}\int_0^\pi &\sin[(m+1)k]\sin[(n+1)k] \nonumber\\
  & \times e^{-uM^2\left(2-2\sin k\right)}\dd k.
  \label{eq:AuM}
\end{align}

\subsection{Continuum operator limit}
For fixed $u>0$, the exponential factor in \eqref{eq:AuM} localizes exponentially near $k=\pi/2$ as $M\to\infty$, which is the first step towards a continuum limit.
The joint scaling limit $M\to\infty$ with fixed $u>0$ yields a compact continuum optimization problem.
To make the scaling explicit, set $k=\pi/2+\xi/M$.
Then
\begin{equation}
uM^2\left(2-2\sin(\tfrac{\pi}{2}+\xi/M)\right)=u\xi^2+O(M^{-2}\xi^4),
\label{eq:weight_expand}
\end{equation}
and $\dd k=\dd \xi/M$.
The Gaussian factor $e^{-u\xi^2}$ therefore restricts the effective integration to $\abs{\xi}=O(1)$.
This near-peak concentration is visible already at finite $M$, and biasing the optimization toward the extreme-velocity tail further sharpens the localization, as illustrated in Fig.~\ref{fig:optbias}(b).

This scaling also isolates the degrees of freedom that remain relevant under the real-space support constraint. As noted in Sec.~\ref{sec:finiteM}, the sine amplitude $c_\psi(k)$ is a finite sine polynomial. Under the rescaling $k=k_0+\xi/M$ with $k_0=\pi/2$, this produces a family of near-peak envelopes 
\begin{equation}
  \widetilde c_{\psi}^{(M)}(\xi)=c_{\psi}\left(k_0+\frac{\xi}{M}\right),
\label{eq:ctilde_def}
\end{equation} 
whose $M\to\infty$ closure forms the Fourier-window subspace introduced below. The continuum limit therefore acts on these near-peak envelopes rather than on microscopic site amplitudes.

Near $k_0$ the sine basis resolves into cosine and sine components depending on site parity.
Using the sine addition formula,
we find, for $j\ge 0$,
\begin{equation}
\sin\left[(2j+1)\left(\frac{\pi}{2}+\frac{\xi}{M}\right)\right]=(-1)^j\cos\left[\frac{(2j+1)\xi}{M}\right],
\label{eq:even_indices}
\end{equation}
and
\begin{equation}
\sin\left[(2j+2)\left(\frac{\pi}{2}+\frac{\xi}{M}\right)\right]=(-1)^{j+1}\sin\left[\frac{(2j+2)\xi}{M}\right].
\label{eq:odd_indices}
\end{equation}
Writing $\alpha_j:=(-1)^j\psi_{2j}$ and $\beta_j:=(-1)^{j+1}\psi_{2j+1}$ gives
\begin{align}
\widetilde c_{\psi}^{(M)}(\xi)=& \sqrt{\frac{2}{\pi}}
\sum_j \alpha_j\cos\left[\frac{(2j+1)\xi}{M}\right]\nonumber\\
&+\sqrt{\frac{2}{\pi}}
\sum_j \beta_j\sin\left[\frac{(2j+2)\xi}{M}\right].
\label{eq:ctilde_cos_sin}
\end{align}

Because the weight in \eqref{eq:AuM} becomes an even function of $\xi$ on a symmetric $\xi$ domain, the cosine-sine cross term in the associated quadratic form vanishes by parity.
Consequently, the near-peak optimization decouples into cosine and sine sectors. For any fixed $u>0$, the continuum operator obtained below commutes with the parity transform $(\Pi f)(\xi)=f(-\xi)$, and its principal eigenfunction is even. Thus the cosine sector gives the leading limiting eigenvalue, while Appendix~\ref{app:optimized_values} keeps both sectors in the convergence proof.

In either sector, the $M$-site constraint turns into a sharp Fourier cutoff $\abs{p}\le 1$ in the Fourier variable $p$ conjugate to $\xi$. We denote by $\mathcal{H}_{\rm W}\subset L^2(\mathbb{R})$ the corresponding Fourier-window subspace, namely the set of functions whose Fourier transform is supported in $[-1,1]$~\cite{SlepianPollak1961,LandauPollak1961,LandauWidom1980TimeFrequency,Widom1964IntegralEquationsII}. We use $\widehat g(p)=\int_{\mathbb{R}}e^{-\ii p \xi}g(\xi)\dd \xi$ and $g(\xi)=\int_{\mathbb{R}}e^{\ii p \xi}\widehat g(p)\dd p /(2\pi)$.
Let $P$ be the orthogonal projection onto $\mathcal{H}_{\rm W}$, which has the sinc kernel
\begin{equation}
(Pf)(\xi)=\int_{\mathbb{R}}\frac{\sin(\xi-\eta)}{\pi(\xi-\eta)}f(\eta)\dd \eta.
\label{eq:projP}
\end{equation}
In Fourier space, $P$ implements the sharp cutoff $\widehat f(p)\mapsto \1_{[-1,1]}(p)\widehat f(p)$ where $\1_{[-1,1]}(p)$ is the indicator function of the interval $[-1,1]$.

Define the Gaussian multiplier $(M_{u/2}f)(\xi)=e^{-u\xi^2/2}f(\xi)$ and the compact self-adjoint operator
\begin{equation}
T_u=M_{u/2}PM_{u/2}.
\label{eq:Tu}
\end{equation}
The Gaussian factors represent the bias toward the extreme-velocity tail through the near-front defect variable, and $T_u$ packages the combined effect of this bias with the sharp Fourier cutoff.
In $\xi$ space, $T_u$ has the Gaussian-windowed sinc kernel
\begin{equation}
(T_uf)(\xi)=\int_{\mathbb{R}}e^{-u(\xi^2+\eta^2)/2}\frac{\sin(\xi-\eta)}{\pi(\xi-\eta)}f(\eta)\dd \eta.
\label{eq:Tu_kernel}
\end{equation}

A useful equivalent representation is obtained by observing that $T_u$ can be factored as $T_u=AA^\dagger$ with $A=M_{u/2}P$.
Hence $T_u$ and the operator $A^\dagger A=P M_u P$ have the same nonzero spectrum.
In particular,
\begin{equation}
\lambda_{\max}(T_u)=\lambda_{\max}(P M_u P).
\label{eq:spectrum_equiv}
\end{equation}
This representation makes the variational structure explicit. For any $g\in\mathcal{H}_{\rm W}$ with $\norm{g}_2=1$,
\begin{equation}
\mel{g}{P M_u P}{g}=\int_{\mathbb{R}}e^{-u \xi^2}\abs{g(\xi)}^2\dd \xi,
\label{eq:var_lambda}
\end{equation}
so $\lambda_{\max}(T_u)$ is the maximal achievable Gaussian-weighted mass in real space among unit-norm functions in $\mathcal{H}_{\rm W}$.

\subsection{Fourier representation, positivity, and scaling function}

In Fourier space the operator becomes positivity preserving.
Let $\widehat g(p)=\int_{\mathbb{R}}e^{-\ii p \xi}g(\xi)\dd \xi$.
For $g\in\mathcal{H}_{\rm W}$, define $\phi(p):=(2\pi)^{-1/2}\widehat g(p)$ on $[-1,1]$, so that $\norm{g}_2=\norm{\phi}_{L^2([-1,1])}$ by the Plancherel theorem, i.e., the unitarity of the Fourier transform on $L^2$.
In this representation $P M_u P$ acts on $\phi\in L^2([-1,1])$ as the integral operator with strictly positive Gaussian kernel
\begin{align}
 &(\mathsf{S}_u\phi)(p)=\int_{-1}^{1}\mathcal{K}_u(p,q)\phi(q)\dd q,\\
 &\mathcal{K}_u(p,q)=\frac{1}{2\pi}\sqrt{\frac{\pi}{u}}
 \exp\left[-\frac{(p-q)^2}{4u}\right].
 \label{eq:Su_kernel}
\end{align}

Consequently, the top eigenvalue is nondegenerate and the corresponding eigenfunction can be chosen strictly positive on $[-1,1]$~\cite{KreinRutman1948}.
This yields uniqueness up to a phase of the optimal near-peak state.
Define the continuum scaling function
\begin{equation}
\lambda(u)=\lambda_{\max}(T_u),\ 
\Phi(u)=\ln\lambda(u).
\label{eq:lambdaPhi}
\end{equation}
The joint scaling limit can be stated as
\begin{equation}
\mathcal{G}_M(u)\xrightarrow[M\to\infty]{}\lambda(u),
\label{eq:scalinglimit}
\end{equation}
for each fixed $u>0$. The endpoint $u=0$ is immediate since both sides are equal to one.

\begin{figure}[!t]
 \centering
 \includegraphics[width=\linewidth]{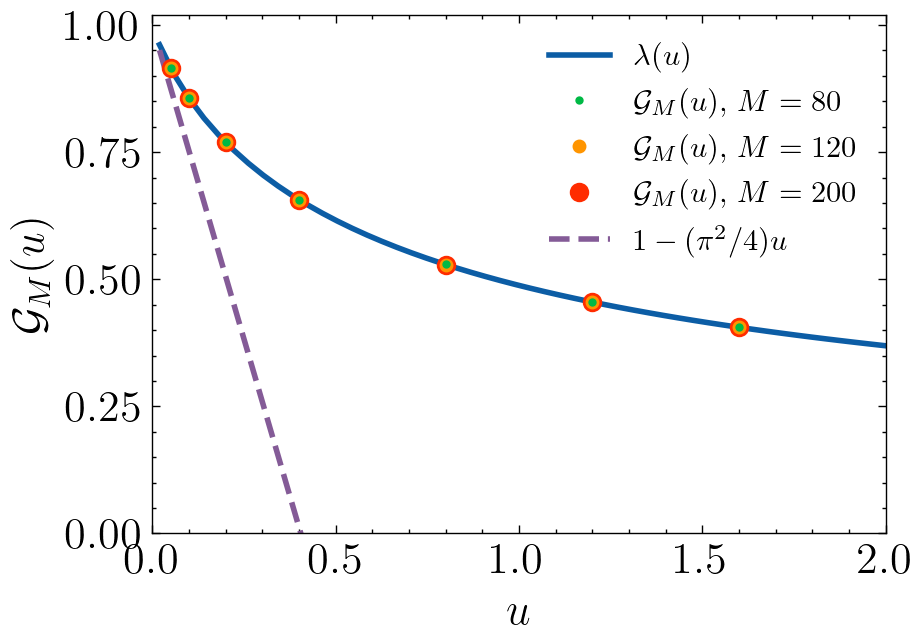}
 \caption{Scaling function for the optimized near-peak statistics. The solid curve is the continuum principal eigenvalue $\lambda(u)=\lambda_{\max}(T_u)$ computed from the Fourier-side kernel \eqref{eq:Su_kernel}. Markers show finite-$M$ values of $\mathcal{G}_M(u)$ obtained by diagonalizing the discrete matrix $A_u^{(M)}$ in \eqref{eq:AuM}. The dashed line is the small-$u$ expansion $1-(\pi^2/4)u$.}
 \label{fig:lambda_u}
\end{figure}

The discussion above isolates the two scaling mechanisms that survive in the continuum limit. The quadratic peak produces the Gaussian window in $\xi$, and the $M$-site support produces the sharp Fourier cutoff $\abs{p}\le 1$.
An explicit Fourier embedding of the discrete cosine and sine series into the Fourier-window subspace is given in Appendix~\ref{app:fourier_embedding}, and the convergence argument is completed in Appendix~\ref{app:optimized_values}.
A numerical illustration of the scaling function $\lambda(u)$ and of the convergence of $\mathcal{G}_M(u)$ is given in Fig.~\ref{fig:lambda_u} in Sec.~\ref{sec:numerics}.

\section{Asymptotics of the maximal mean drift}
\label{sec:constant}
\subsection{From defect statistics to optimal mean drift}

Because $\Delta_M=M^2(2J-V)/J$, the mean drift can be recovered from the defect:
\begin{equation}
\E_\psi[V]=2J-\frac{J}{M^2}\E_\psi[\Delta_M].
\end{equation}
In the optimized setting, define
\begin{equation}
\Phi_M(u)=\ln \mathcal{G}_M(u).
\label{eq:PhiM}
\end{equation}
For $u>0$, the maximizer typically depends on $u$, so derivatives of $\Phi_M$ describe how the $u$-optimal family responds when the bias is increased.
At $u=0$ the maximization is degenerate since $\mathcal{G}_M(0)=1$ for every $\psi$, and the right derivative at zero bias has a direct variational meaning as the optimal mean defect.
In particular, we do not interpret $\Phi_M''(0^+)$ as an optimization of the variance because the maximizing state typically depends on $u$.

At fixed finite $M$, this endpoint derivative can be obtained directly without differentiating through the degenerate set of maximizers.
Let $D^{(M)}$ be the finite-dimensional defect matrix
\begin{align}
& (D^{(M)})_{mn}\nonumber\\
& =\frac{2}{\pi}\int_0^\pi \sin[(m+1)k]\sin[(n+1)k] M^2(2-2\sin k)\dd k.
\label{eq:defect_matrix}
\end{align}
Since $A_u^{(M)}=I_M-uD^{(M)}+O_M(u^2)$ in operator norm as $u\downarrow0$, Rayleigh-Ritz gives
\begin{equation}
\mathcal{G}_M(u)=1-u\lambda_{\min}(D^{(M)})+O_M(u^2).
\label{eq:GM_endpoint_expansion}
\end{equation}
Define the optimal mean defect
\begin{equation}
d_M:=\min_{\psi\in\mathcal{S}_M}\E_\psi[\Delta_M]=\lambda_{\min}(D^{(M)})=-\Phi_M'(0^+).
\label{eq:dM_def}
\end{equation}
The operator-norm expansion leading to Eq.~\eqref{eq:GM_endpoint_expansion} is given in Appendix~\ref{app:finiteM_endpoint}.
By definition \eqref{eq:vmax_def}, we have the exact relation
\begin{equation}
2J-v_{\max}(M)=\frac{J}{M^2}d_M.
\label{eq:vmax_from_PhiM}
\end{equation}

\begin{figure}[!t]
 \centering
 \includegraphics[width=\linewidth]{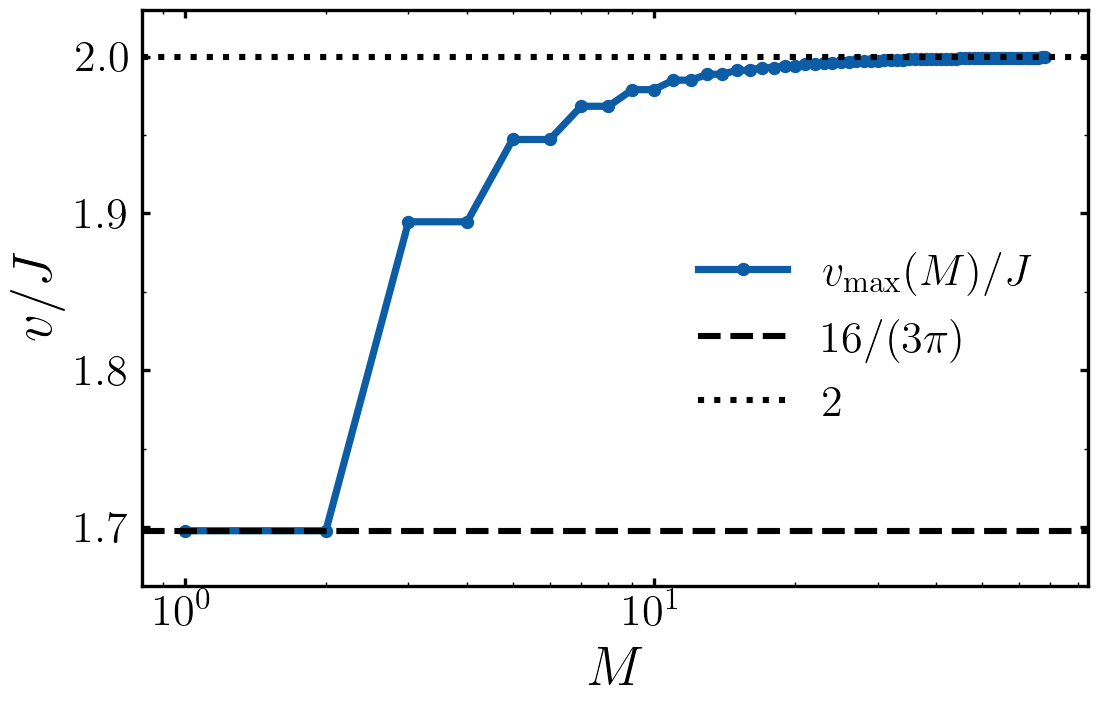}
 \caption{Coherent optimum versus the site-diagonal incoherent reference for $J=1$. The solid line is $v_{\max}(M)/J$ obtained by diagonalizing $A^{(M)}$. The dashed horizontal line is $v_{\rm inc}^{\max}/J=16/(3\pi)$ from Eq.~\eqref{eq:incoherent_reference}. The dotted horizontal line marks the ballistic front $2J/J=2$.}
 \label{fig:coherence}
\end{figure}

To extract the large-$M$ asymptotics, it is not necessary to interchange the limit $M\to\infty$ with the right derivative at $u=0$.
It is enough to combine the fixed-$u$ scaling limit \eqref{eq:scalinglimit} with a lower bound on $d_M$ that holds for every $u>0$.
For any fixed $\psi$ and any $u>0$, Jensen's inequality gives
\begin{equation}
-\frac{1}{u}\ln \E_\psi[e^{-u\Delta_M}] \le \E_\psi[\Delta_M].
\end{equation}
Let $\psi\in\arg\min_{\phi\in\mathcal{S}_M}\E_\phi[\Delta_M]$.
Applying Jensen's inequality to this choice and using $\E_\psi[e^{-u\Delta_M}]\le \mathcal{G}_M(u)=e^{\Phi_M(u)}$ gives
\begin{equation}
-\frac{1}{u}\Phi_M(u)\le d_M.
\label{eq:dM_lower_u}
\end{equation}
For each fixed $u>0$, the scaling limit \eqref{eq:scalinglimit} implies $\Phi_M(u)\to \Phi(u)=\ln\lambda(u)$, hence
\begin{equation}
\liminf_{M\to\infty} d_M \ge -\frac{1}{u}\Phi(u).
\label{eq:liminf_dM_u}
\end{equation}
See Appendix~\ref{app:proof_liminf_dM_u} for a direct proof.
Since $\lambda(0)=1$, we have $\Phi(0)=0$.
We define
\begin{equation}
C_1=-\Phi'(0^+)=-\lim_{u\to 0^+}\frac{\Phi(u)}{u}.
\label{eq:def_C1}
\end{equation}
Appendix~\ref{app:proof_varC1} proves directly, without max-differentiation at the degenerate endpoint, that
\begin{equation}
\lambda(u)=1-\frac{\pi^2}{4}u+o(u),
\qquad C_1=\frac{\pi^2}{4}.
\label{eq:lambda_smallu_main}
\end{equation}
Therefore, inequality~\eqref{eq:liminf_dM_u} yields
\begin{equation}
\liminf_{M\to\infty} d_M \ge C_1.
\label{eq:liminf_dM}
\end{equation}

Conversely, fix a real even profile $g\in\mathcal{H}_{\rm W}$ with $\norm{g}_2=1$, finite second moment, and a smooth Fourier amplitude compactly supported in $(-1,1)$.
Using the explicit embedding in Appendix~\ref{app:proof_trial_mean_defect}, we can construct a sequence of states $\psi^{(M)}\in\mathcal{S}_M$ whose embedded near-peak profiles approximate $g$ and such that the corresponding mean defects satisfy
\begin{equation}
\lim_{M\to\infty}\E_{\psi^{(M)}}[\Delta_M]=\int_{\mathbb{R}}\xi^2\abs{g(\xi)}^2\dd \xi.
\label{eq:trial_mean_defect}
\end{equation}
Since $d_M$ is the minimum of $\E_\psi[\Delta_M]$ over $\mathcal{S}_M$, this gives
\begin{equation}
\limsup_{M\to\infty} d_M \le \int_{\mathbb{R}}\xi^2\abs{g(\xi)}^2\dd \xi.
\end{equation}
Such smooth even Fourier profiles are dense in the even part of $H_0^1((-1,1))$, which contains the Dirichlet ground-state minimizer in Eq.~\eqref{eq:varC1}.
Optimizing over all such $g$ yields
\begin{equation}
\limsup_{M\to\infty} d_M \le C_1.
\label{eq:limsup_dM}
\end{equation}
Combining \eqref{eq:liminf_dM} and \eqref{eq:limsup_dM} shows that $d_M=C_1+o(1)$ as $M\to\infty$.
Substituting this into the exact relation \eqref{eq:vmax_from_PhiM} yields
\begin{equation}
v_{\max}(M)=2J-\frac{J}{M^2}C_1+o(M^{-2}).
\label{eq:vmax_asym_C1}
\end{equation}
Using $C_1=\pi^2/4$ from Eq.~\eqref{eq:lambda_smallu_main} gives
\begin{equation}
v_{\max}(M)=2J-\frac{\pi^2J}{4M^2}+o(M^{-2}).
\label{eq:vmax_final}
\end{equation}
Thus the main asymptotic result follows. The remaining part of this section gives the Fourier-side interpretation of the coefficient appearing in Eq.~\eqref{eq:vmax_final}.
\subsection{Variational origin of the coefficient}
The argument in the previous subsection fixes the coefficient through endpoint bounds. Its value has a simple variational origin in the continuum Fourier-window problem.
In Fourier space, multiplication by $\xi$ corresponds to $\ii\partial_p$.
If $g$ is generated by a Fourier-side function $\phi(p)=(2\pi)^{-1/2}\widehat g(p)$ supported in $[-1,1]$, then
\begin{equation}
\int_{\mathbb{R}}\xi^2\abs{g(\xi)}^2\dd \xi=\int_{-1}^{1}\abs{\phi'(p)}^2\dd p,
\label{eq:x2toDeriv_main}
\end{equation}
provided $\phi$ has a square-integrable weak derivative and vanishes at $p=\pm 1$ in the trace sense.
This regularity is the Fourier-side expression of a finite second moment. If the Fourier profile does not vanish at the cutoff points $p=\pm 1$, the sharp Fourier cutoff produces a slow real-space tail and the second moment diverges~\cite{LandauWidom1980TimeFrequency,Widom1964IntegralEquationsII,SlepianPollak1961,LandauPollak1961}.
In this representation, the finite second moment selects profiles whose Fourier amplitude is sufficiently smooth and vanishes at the cutoff points, which is the Dirichlet condition for the associated variational problem.
The dimensionless coefficient is therefore the Rayleigh quotient of the Dirichlet Laplacian on $[-1,1]$:
\begin{equation}
\inf\left\{\frac{\int_{-1}^{1}\abs{f'(p)}^2 \dd p}{\int_{-1}^{1}\abs{f(p)}^2 \dd p}: f\in H_0^1((-1,1)),  f\not\equiv 0\right\}.
\label{eq:varC1}
\end{equation}
By the Rayleigh-Ritz principle \cite{Ritz1909,MacDonald1933RayleighRitz}, the infimum equals the lowest eigenvalue of $- \partial_p^2$ on $[-1,1]$ with Dirichlet boundary conditions.
Solving $-f''=\mu f$ with $f(\pm 1)=0$ gives $\mu_1=(\pi/2)^2$, attained for example by $f(p)=\cos(\pi p/2)$.
This is the variational origin of the coefficient $\pi^2/4$ in Eq.~\eqref{eq:vmax_final}.
The value of this dimensionless coefficient is tied to the nearest-neighbor hard-wall model studied here.
It combines the quadratic maximum of the group velocity at $k_0=\pi/2$ with the sharp Fourier cutoff produced by the $M$-site support constraint.
For other dispersions or for different preparation constraints, the same scaling mechanism may lead to a different continuum variational problem and hence to a different numerical coefficient.

\begin{figure}[!t]
 \centering
 \includegraphics[width=\linewidth]{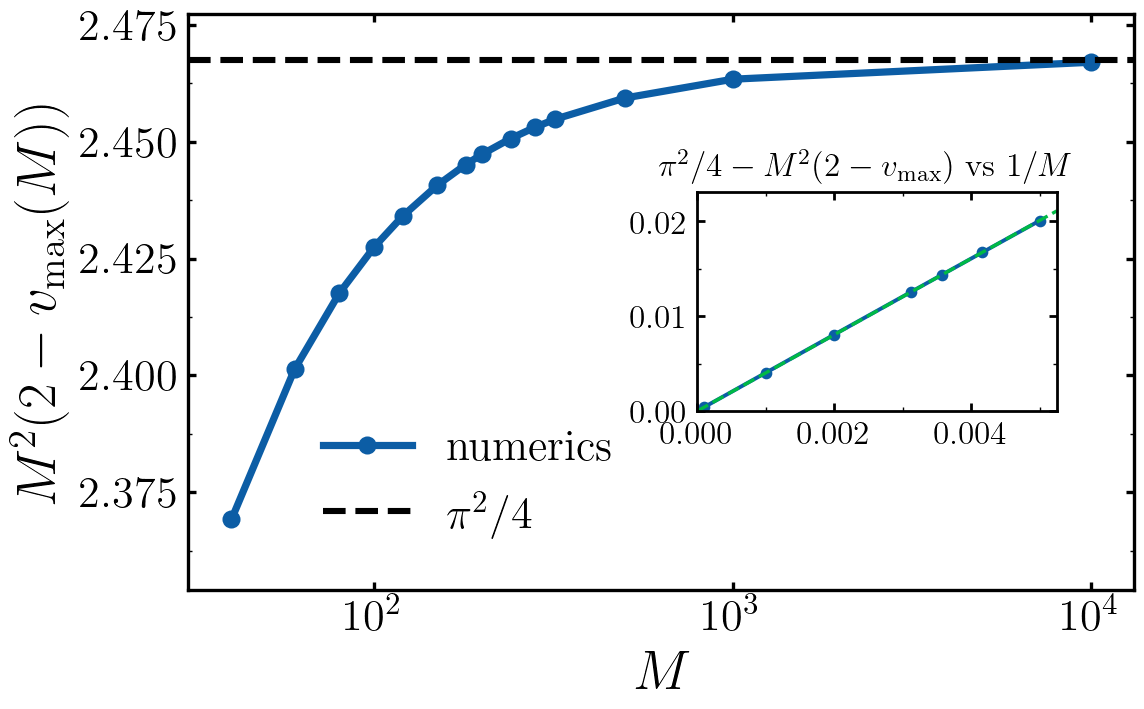}
 \caption{Optimized mean drift for $J=1$. The main panel shows the rescaled estimator $M^2(2-v_{\max}(M))$ versus $M$ obtained from diagonalizing $A^{(M)}$. The dashed line is the constant limit $\pi^2/4$. The inset shows the residual $\pi^2/4-M^2(2-v_{\max}(M))$ versus $1/M$.}
 \label{fig:vmax}
\end{figure}

\section{Numerical illustration and finite-$M$ convergence}
\label{sec:numerics}
The vital point of the spectral formulation is that all quantities of interest reduce to eigenvalues and eigenvectors of explicit finite-dimensional matrices.
All numerical results in this work are obtained by direct diagonalization of $A^{(M)}$ and $A_u^{(M)}$ defined in Eqs.~\eqref{eq:Adef} and \eqref{eq:AuM}, or of the continuum operator $T_u$ in Eq.~\eqref{eq:Tu}.
No real-time evolution is required.
Implementation details and convergence checks are summarized in Appendix~\ref{app:numerics}.

The optimal finite-$M$ initial states and their near-peak spectral concentration are illustrated in Fig.~\ref{fig:optbias}.
For the mean-drift optimum, the real-space structure is parity selected: the maximizing eigenvector can be chosen in the even-site block of $A^{(M)}$, so the odd-site amplitudes vanish.
On this sublattice, Appendix~\ref{app:proof_trial_mean_defect} provides an asymptotically optimal large-$M$ trial family, $\psi_{2j}^{(M,0)}=(-1)^j\alpha_j^{(M)}$ and $\psi_{2j+1}^{(M,0)}=0$, which captures the slowly varying envelope and alternating signs of the leading finite-support structure.
The same parity selection holds for the defect-biased matrices $A_u^{(M)}$ for every fixed $u>0$, so the $u=1$, $M=200$ optimizer in Fig.~\ref{fig:optbias} has the same support pattern.

Figure~\ref{fig:lambda_u} illustrates the scaling function $\lambda(u)$ together with finite-$M$ values of $\mathcal{G}_M(u)$ obtained from the discrete eigenproblem \eqref{eq:GM_eig}.
The continuum curve is computed by discretizing the compact operator $T_u$ in Fourier representation \eqref{eq:Su_kernel}, while the markers come from diagonalizing $A_u^{(M)}$ for increasing $M$. 
The agreement is already accurate for moderate $M$, reflecting the fact that the optimization localizes spectral weight near $k=\pi/2$ and suppresses sensitivity to the sine polynomial away from the peak.

Figure~\ref{fig:coherence} compares the coherent optimum with the site-diagonal reference value \eqref{eq:incoherent_reference}.
The reference value is reached already by the boundary-localized state and does not improve with $M$, whereas the coherent optimum increases with support size and approaches the ballistic front.
This comparison illustrates that the inverse-square approach to $2J$ is a coherent finite-support effect rather than an optimization over occupations alone.

The same even--odd block structure also produces exact pairwise plateaus in the coherent optimum,
\begin{equation}
v_{\max}(2r-1)=v_{\max}(2r),\quad r\ge 1.
\label{eq:pairwise_plateau_main}
\end{equation}
As proved in Appendix~\ref{app:parity_plateau}, the even-site block carries the larger principal eigenvalue of $A^{(M)}$.
Adding site $2r-1$ only enlarges the odd-site block and does not change the optimum, whereas adding site $2r$ enlarges the even-site block and raises $v_{\max}$.

Figure~\ref{fig:vmax} tests the asymptotic law \eqref{eq:vmax_final} for $J=1$ using direct diagonalization of $A^{(M)}$.
The main panel plots the rescaled estimator $M^2(2-v_{\max}(M))$ as a function of $M$, which approaches the constant limit $\pi^2/4$.
The inset shows the residual $\pi^2/4-M^2(2-v_{\max}(M))$ versus $1/M$ and is consistent with an $O(1/M)$ correction on the accessible sizes.

\section{Experimental relevance}
\label{sec:exp_relevance}

Our setup is naturally compatible with ultracold-atom systems where single-particle tunneling in a one-dimensional optical lattice realizes a continuous-time quantum walk generated by a nearest-neighbor tight-binding Hamiltonian. Quantum walk transport and its spatiotemporal spreading have been observed in optical lattices, and quantum gas microscopes provide site-resolved imaging that turns the full real space evolution into a directly measurable observable \cite{Preiss2015StronglyCorrelatedQW,GrossBakr2021QGM}.

The hard-wall half-line can be approximated by a steep repulsive barrier at the origin, provided tunneling across the barrier is negligible on the observation timescale \cite{LiuNakagawaUeda2023MaxwellDemonTransport}. A temporary preparation potential can localize a single atom predominantly on the first $M$ sites. Removing this potential at $t=0$ while retaining the origin barrier leaves the subsequent evolution governed by the half-line Hamiltonian. Quantum-gas microscopes provide the site-resolved control of local potentials and barrier positions required for this prepare-and-release protocol \cite{GrossBakr2021QGM,DiCarli2024CommensurateIncommensurate}.

Experimental preparations generally leave a small residual population outside the target region. The ideal hard-support problem therefore provides a reference for preparations with small leakage. In a lattice microscope, shaped local potentials can concentrate the initial wavefunction on the first $M$ sites before the quench. Neutral-atom tweezer arrays provide a complementary route through deterministic assembly and placement on selected lattice sites \cite{Kim2016HolographicTweezers}.

The exact optimal state is the principal eigenvector of the $M$-dimensional matrix $A^{(M)}$ and defines the optimal reference value under the preparation constraint. For the mean-drift optimum, Appendix~\ref{app:parity_plateau} shows that the maximizing eigenvector can be chosen on the even sublattice. Appendix~\ref{app:proof_trial_mean_defect} gives an asymptotically optimal large-$M$ trial family on this sublattice, with a slowly varying envelope and alternating signs. Simpler coherent state families can test the finite-support dependence by comparing their measured drift with the optimal value as $M$ is varied.

The mean ballistic drift can be extracted from the large-time growth of the mean position and from the convergence of $X_t/t$ to a stationary velocity law. Repeating the preparation with different target sizes $M$ varies the initial support without changing the post-quench half-line Hamiltonian. Measurements for a sequence of target sizes $M$ can test the predicted $M^{-2}$ deficit using wavepackets with quantified leakage outside the target region. The measured drift can then be compared with the optimal prefactor derived here.
 
\section{Discussion and outlook}
\label{sec:discussion}

The hard wall does not destroy ballisticity but changes how initial-state resources determine how closely one can approach the velocity peak.
Under a finite-support resource constraint, the half-line CTQW Hamiltonian \eqref{eq:H} is unitarily equivalent to the restriction of an open XX chain to its single-excitation sector \cite{Bose2003SpinChainComm,Christandl2004PerfectTransfer,Kay2010PerfectTransferReview}.
This connection places our ballistic front in the broader language of quantum many-body locality. Lieb-Robinson bounds \cite{LiebRobinson1972FiniteVelocity,NachtergaeleSims2006LRClustering,HastingsKoma2006LRDecay} imply an effective light cone in local spin systems, and in the present single-excitation setting the front is set by the maximal group velocity of the one-particle dispersion.
Our main result establishes that under the finite-support initial-state constraint, the optimal mean drift satisfies $2J-v_{\max}(M)=\pi^2 J/(4M^2)+o(M^{-2})$.

The weak-gradient saturation of the drift reported in Maxwell-demon-assisted transport \cite{LiuNakagawaUeda2023MaxwellDemonTransport} is naturally interpreted as the hard-wall mean drift of a boundary-localized state, while our results characterize the optimal statistics achievable under broader finite-support resources.
More broadly, recent progress on quantum batteries and closed-loop quantum control has enabled experimentally accessible protocols where state preparation and real-time control directly shape charging and transport dynamics \cite{Campaioli2024QuantumBatteriesColloquium,Yu2024MaxwellDemonBattery,CampagneIbarcq2013PersistentControl,Doherty2000QuantumFeedbackControl}.
On the other hand, topological classification of quantum feedback control is discussed in recent work \cite{NakagawaUeda2025TopologyFeedback} and demon-assisted chiral transport on one-dimensional lattice is also proposed.
These developments motivate quantifying how finite state-preparation resources bound the best achievable approach to a ballistic front, for which the present finite-support formulation provides an analytically tractable reference point.

The near-peak scaling function $\lambda(u)$ contains substantially more information than the single coefficient $\pi^2/4$.
For $u>0$, increasing the bias selects a distinguished $u$-dependent family of optimal states.
In this sense, $\Phi_M(u)=\ln\mathcal{G}_M(u)$ can be viewed, for $u>0$, as the cumulant generating function of $\Delta_M$ evaluated in that $u$-optimal family \cite{Touchette2009LargeDeviationsReview,ChetriteTouchette2015Conditioned}.
Formally, if higher right derivatives of $\Phi$ at $u=0^+$ exist and if differentiation commutes with the scaling limit, then the coefficients
\begin{equation}
\kappa_m(\Delta)=(-1)^m\Phi^{(m)}(0^+)
\label{eq:cumulants}
\end{equation}
govern the small-$u$ expansion of $\Phi_M(u)$ and would imply the hierarchy $\kappa_m(2J-V)\sim J^m M^{-2m}$ (equivalently, $\kappa_m(\Delta_M)=O(1)$) for the infinitesimal-tilt-selected family.
We do not interpret these coefficients as cumulants that are separately optimized at fixed order, because the optimizing state typically depends on $u$ and the envelope at $u=0$ requires a careful branch-selection analysis.
Therefore we treat Eq.~\eqref{eq:cumulants} as a conditional statement unless additional regularity is assumed.

Even without a full analytic theory of all derivatives at $u=0$, the function $\lambda(u)$ itself is an observable scaling object.
For $u>0$, $\lambda(u)$ is smooth and can be computed reliably from the Fourier representation \eqref{eq:Su_kernel}.
In this sense, $\lambda(u)$ provides a description of how the best achievable near-peak statistics depends on the bias strength $u$, interpolating between the unbiased point $\lambda(0)=1$ and strongly biased regimes where only extremely narrow near-peak profiles contribute. We leave this higher-order analysis for future work.

\begin{acknowledgments}
This work was supported by Tianfu talent project. K. L. was supported by the Scientific Research Startup
Foundation of Xihua University (Grant No: Z241064).
\end{acknowledgments}

\appendix

\section{Discrete-to-continuum scaling and the operator $T_u$}
\label{app:continuum}

\subsection{Proof of Eq.~\eqref{eq:scalinglimit}}
\label{app:proof_scalinglimit}

\subsubsection{Band limitation and an explicit Fourier embedding}
\label{app:fourier_embedding}

To connect \eqref{eq:ctilde_cos_sin} to the Fourier-window subspace, it is convenient to pass from cosines and sines to complex exponentials.
For instance,
\begin{equation}
\cos\left[\frac{(2j+1)\xi}{M}\right]=\frac{1}{2}\left(e^{\ii p_j \xi}+e^{-\ii p_j \xi}\right),
\end{equation}
with $p_j:=(2j+1)/M\in(0,1]$ and similarly for the sine series.
A concrete embedding is obtained by defining a Fourier profile on $[-1,1]$ that approximates these discrete frequencies.
For notational simplicity, we describe the cosine even-index sector. The sine sector is analogous.

Let $h_M:=2/M$, $N_{\rm c}:=\lceil M/2\rceil$, and $\alpha_j:=(-1)^j\psi_{2j}$ for $j=0,\dots,N_{\rm c}-1$.
Define the truncated cells
\begin{equation}
I_j^+:=\left[p_j-\frac{h_M}{2},p_j+\frac{h_M}{2}\right]\cap[0,1],
\quad I_j^-:=-I_j^+,
\end{equation}
and a step-function Fourier profile
\begin{equation}
\widehat g_M(p):=\sum_{j=0}^{N_{\rm c}-1}\sqrt{\frac{\pi}{|I_j^+|}}\alpha_j
\left(\1_{I_j^+}(p)+\1_{I_j^-}(p)\right),
\label{eq:gM_hat}
\end{equation}
where $p\in[-1,1]$.
Then $\widehat g_M$ is supported in $[-1,1]$.
Define $g_M$ by inverse Fourier transform with the convention stated in Sec.~\ref{sec:continuum}:
\begin{equation}
g_M(\xi)=\frac{1}{2\pi}\int_{-1}^{1}e^{\ii p \xi}\widehat g_M(p)\dd p.
\label{eq:gM_def}
\end{equation}
By construction $g_M\in\mathcal{H}_{\rm W}$.
Moreover, since $\widehat g_M$ is piecewise constant, $g_M$ is essentially a superposition of sinc-like wave packets centered at the discrete frequencies.

A direct computation from \eqref{eq:gM_hat} gives
\begin{equation}
\label{eq:gM_norm}
\norm{\widehat g_M}_2^2
=
2\pi\sum_{j=0}^{N_{\rm c}-1}\abs{\alpha_j}^2,
\end{equation}
and by the Plancherel theorem $\norm{g_M}_2^2=\sum_j\abs{\alpha_j}^2$.
This embedding identifies the limiting Fourier window. The uniform passage to optimized values is established below in both parity sectors.

\subsubsection{Emergence of the Gaussian kernel in frequency space}

The key step in the proof of Eq.~\eqref{eq:scalinglimit} is that the Gaussian weight turns the $\xi$-space quadratic form into a Gaussian kernel in frequency space.
Let $g(\xi)=\int_{-1}^{1}e^{\ii p \xi}\widehat g(p)\dd p /(2\pi)$ be any function in $\mathcal{H}_{\rm W}$.
Then
\begin{align}
\int_{\mathbb{R}} e^{-u \xi^2}\abs{g(\xi)}^2\dd \xi
&=\int_{\mathbb{R}} e^{-u \xi^2}\left(\frac{1}{2\pi}\int_{-1}^{1}e^{\ii p \xi}\widehat g(p)\dd p\right)\nonumber\\
&\quad \times \left(\frac{1}{2\pi}\int_{-1}^{1}e^{-\ii q \xi}\overline{\widehat g(q)}\dd q\right)\dd \xi \nonumber\\
&=\frac{1}{(2\pi)^2}\int_{-1}^{1}\int_{-1}^{1}\overline{\widehat g(q)}\widehat g(p)\nonumber\\
&\quad \times \left(\int_{\mathbb{R}} e^{-u \xi^2}e^{\ii(p-q)\xi}\dd \xi\right)\dd p\dd q.
\label{eq:Gaussian_kernel_derivation}
\end{align}
The inner integral is explicit:
\begin{equation}
\int_{\mathbb{R}} e^{-u \xi^2}e^{\ii(p-q)\xi}\dd \xi
=\sqrt{\frac{\pi}{u}}\exp\left[-\frac{(p-q)^2}{4u}\right].
\end{equation}
Substituting into \eqref{eq:Gaussian_kernel_derivation} yields
\begin{equation}
\int_{\mathbb{R}} e^{-u \xi^2}\abs{g(\xi)}^2\dd \xi
=\frac{1}{2\pi}\int_{-1}^{1}\int_{-1}^{1}\overline{\widehat g(q)}\widehat g(p)
\mathcal{K}_u(p,q)\dd p\dd q,
\end{equation}
with the kernel $\mathcal{K}_u(p,q)$ in \eqref{eq:Su_kernel}.
Thus the limiting optimization is exactly a principal eigenvalue problem for a compact integral operator with a strictly positive Gaussian kernel in frequency space.

\subsubsection{Convergence of the optimized values}
\label{app:optimized_values}
The convergence $\mathcal{G}_M(u)\to\lambda(u)$ for fixed $u>0$ follows from operator-norm convergence under an isometric Fourier-cell embedding.
Using the Rayleigh-Ritz reduction in Eq.~\eqref{eq:GM_eig}, we may write
\begin{equation}
\mathcal{G}_M(u)=\max_{\psi\in\mathcal{S}_M,\ \norm{\psi}=1}\expval{A_u^{(M)}}{\psi}.
\label{eq:GM_RR_app}
\end{equation}
Set
\begin{align}
D_M&:=\left[-\frac{\pi M}{2},\frac{\pi M}{2}\right],\\
W_M^{(u)}(\xi)&:=\1_{D_M}(\xi)
\exp\left[-uM^2\left(2-2\cos\frac{\xi}{M}\right)\right].
\label{eq:WM_def}
\end{align}
Let $N_{\rm s}:=\lfloor M/2\rfloor$, and define
\begin{align}
p_j^{\rm c}&:=\frac{2j+1}{M}, & p_j^{\rm s}&:=\frac{2j+2}{M},\\
\alpha_j&:=(-1)^j\psi_{2j}, & \beta_j&:=(-1)^{j+1}\psi_{2j+1}.
\end{align}
The sine identities at $k=\pi/2+\xi/M$ and parity on $D_M$ give the exact unitary decomposition
\begin{equation}
A_u^{(M)}\simeq A_{u,{\rm c}}^{(M)}\oplus A_{u,{\rm s}}^{(M)},
\label{eq:block_directsum}
\end{equation}
where
\begin{align}
\mel{\alpha}{A_{u,{\rm c}}^{(M)}}{\alpha}
&=\frac{2}{\pi M}\int_{\mathbb R}W_M^{(u)}(\xi)
\left|\sum_{j=0}^{N_{\rm c}-1}\alpha_j\cos(p_j^{\rm c}\xi)\right|^2\dd\xi,\\
\mel{\beta}{A_{u,{\rm s}}^{(M)}}{\beta}
&=\frac{2}{\pi M}\int_{\mathbb R}W_M^{(u)}(\xi)
\left|\sum_{j=0}^{N_{\rm s}-1}\beta_j\sin(p_j^{\rm s}\xi)\right|^2\dd\xi.
\end{align}
For every fixed $u>0$,
\begin{equation}
\norm{W_M^{(u)}-e^{-u\xi^2}}_{L^1(\mathbb R)}\longrightarrow0.
\label{eq:L1_weight}
\end{equation}
Indeed, $1-\cos x\ge2x^2/\pi^2$ for $|x|\le\pi/2$, so $0\le W_M^{(u)}(\xi)\le e^{-4u\xi^2/\pi^2}$ on $D_M$, and dominated convergence applies.
Let $B_{u,\sigma}^{(M)}$ be obtained from $A_{u,\sigma}^{(M)}$ by replacing $W_M^{(u)}$ with $e^{-u\xi^2}$, for $\sigma\in\{{\rm c},{\rm s}\}$.
Every entry of $A_{u,\sigma}^{(M)}-B_{u,\sigma}^{(M)}$ is bounded by $2\delta_M/(\pi M)$, where $\delta_M$ is the norm in \eqref{eq:L1_weight}; since $N_\sigma=O(M)$, the Hermitian row-sum bound gives
\begin{equation}
\norm{A_{u,\sigma}^{(M)}-B_{u,\sigma}^{(M)}}_{\opn}\longrightarrow0.
\label{eq:A_B_op}
\end{equation}
Product-to-sum and the Gaussian Fourier integral yield
\begin{equation}
(B_{u,\sigma}^{(M)})_{jl}
=h_M\left[\mathcal K_u(p_j^\sigma,p_l^\sigma)
+\varepsilon_\sigma\mathcal K_u(p_j^\sigma,-p_l^\sigma)\right],
\label{eq:B_entries}
\end{equation}
where $\varepsilon_{\rm c}=1$ and $\varepsilon_{\rm s}=-1$.
Thus $B_{u,\sigma}^{(M)}$ is the midpoint-rule Nystr\"om discretization of the corresponding parity sector of $\mathsf S_u$~\cite{Nystrom1930IntegralEquations}.

Let
\begin{equation}
I_{j,\sigma}^{+}:=\left[p_j^\sigma-\frac{h_M}{2},p_j^\sigma+\frac{h_M}{2}\right],
\quad I_{j,\sigma}^{-}:=-I_{j,\sigma}^{+},
\end{equation}
and define the orthonormal functions
\begin{equation}
e_{j,\sigma}^{(M)}(p):=\frac{1}{\sqrt{2h_M}}
\left(\1_{I_{j,\sigma}^{+}}(p)+\varepsilon_\sigma\1_{I_{j,\sigma}^{-}}(p)\right).
\label{eq:cell_basis}
\end{equation}
Let $U_{M,\sigma}:\mathbb C^{N_\sigma}\to L^2(\mathbb R)$ map the coordinate basis to \eqref{eq:cell_basis}, and put $Q_{M,\sigma}:=U_{M,\sigma}U_{M,\sigma}^*$.
Write
\begin{equation}
(\mathsf S_u^{\mathbb R}f)(p)=\int_{\mathbb R}\mathcal K_u(p,q)f(q)\dd q.
\end{equation}
The entries of $Q_{M,\sigma}\mathsf S_u^{\mathbb R}Q_{M,\sigma}$ are cell averages of $\mathcal K_u$.
Let $\omega_u(\delta)$ denote the modulus of continuity of $\mathcal K_u$ on $[-2,2]^2$.
The difference between the corresponding cell-average and midpoint matrix entries is bounded by $C h_M\omega_u(C h_M)$, where $C$ is independent of $M$ and of the matrix indices.
Since $N_\sigma h_M$ remains bounded, the Hermitian row-sum estimate gives
\begin{equation}
\norm{U_{M,\sigma}B_{u,\sigma}^{(M)}U_{M,\sigma}^*
-Q_{M,\sigma}\mathsf S_u^{\mathbb R}Q_{M,\sigma}}_{\opn}\longrightarrow0.
\label{eq:B_QSQ}
\end{equation}
Let $R_{\rm c}$ and $R_{\rm s}$ be the orthogonal projections onto even and odd functions supported in $[-1,1]$.
The union of the cells differs from $[-1,1]$ by a set of measure $O(h_M)$, and the cell mesh tends to zero. Density of continuous functions in $L^2([-1,1])$ then gives $Q_{M,\sigma}\to R_\sigma$ strongly on $L^2(\mathbb R)$.
Choose a compactly supported smooth cutoff $\chi$ equal to one on a neighborhood of $[-1,1]$ containing the cell ranges for all large $M$.
The operator $K_\chi:=\chi\mathsf S_u^{\mathbb R}\chi$ is Hilbert-Schmidt.
For a compact operator $K$, strong convergence of the orthogonal projections implies
\begin{equation*}
\norm{(Q_{M,\sigma}-R_\sigma)K}_{\opn}
+\norm{K(Q_{M,\sigma}-R_\sigma)}_{\opn}\longrightarrow0.
\end{equation*}
Applying this compactness lemma to $K_\chi$ gives
\begin{equation}
\norm{Q_{M,\sigma}\mathsf S_u^{\mathbb R}Q_{M,\sigma}
-R_\sigma\mathsf S_u^{\mathbb R}R_\sigma}_{\opn}\longrightarrow0.
\label{eq:QSQ_RSR}
\end{equation}
Combining \eqref{eq:A_B_op}, \eqref{eq:B_QSQ}, and \eqref{eq:QSQ_RSR} yields
\begin{equation}
\norm{U_{M,\sigma}A_{u,\sigma}^{(M)}U_{M,\sigma}^*
-R_\sigma\mathsf S_u^{\mathbb R}R_\sigma}_{\opn}\longrightarrow0.
\label{eq:sector_op_conv}
\end{equation}
The min-max principle gives convergence of the largest eigenvalues in each sector.
The Gaussian kernel on $[-1,1]$ is strictly positive and reflection invariant; by the Krein-Rutman theorem its principal eigenfunction is strictly positive, simple, and even~\cite{Jentzsch1912PositiveKernel,KreinRutman1948}.
Using the exact direct sum \eqref{eq:block_directsum}, we obtain
\begin{equation}
\mathcal G_M(u)\longrightarrow\lambda(u)
\end{equation}
for every fixed $u>0$.
The case $u=0$ is immediate since $\mathcal{G}_M(0)=1$ and $\lambda(0)=1$.

\subsection{Endpoint derivative at finite $M$}
\label{app:finiteM_endpoint}

We prove Eq.~\eqref{eq:GM_endpoint_expansion}.
Since $M$ is fixed, Taylor expansion of the exponential in \eqref{eq:AuM} gives
\begin{equation}
A_u^{(M)}=I_M-uD^{(M)}+O_M(u^2)
\end{equation}
in operator norm as $u\downarrow0$.
By the Rayleigh-Ritz formula,
\begin{align}
\mathcal G_M(u)
&=\max_{\norm{\psi}=1}\mel{\psi}{A_u^{(M)}}{\psi}\nonumber\\
&=1-u\min_{\norm{\psi}=1}\mel{\psi}{D^{(M)}}{\psi}+O_M(u^2)\nonumber\\
&=1-ud_M+O_M(u^2).
\end{align}
Taking logarithms gives $\Phi_M(u)=-ud_M+O_M(u^2)$ and therefore $d_M=-\Phi_M'(0^+)$.

\subsection{Proof of inequality~\eqref{eq:liminf_dM_u}}
\label{app:proof_liminf_dM_u}

Fix $u>0$ and $\psi\in\mathcal{S}_M$.
Since $e^{-ux}$ is convex on $[0,\infty)$, Jensen's inequality gives
\begin{equation}
\E_\psi[e^{-u\Delta_M}]\ge e^{-u\,\E_\psi[\Delta_M]}.
\end{equation}
Taking logarithms and rearranging yields
\begin{equation}
\E_\psi[\Delta_M]\ge -\frac{1}{u}\ln \E_\psi[e^{-u\Delta_M}].
\end{equation}
Choose $\psi$ to minimize $\E_\psi[\Delta_M]$ over $\mathcal S_M$ and use $\E_\psi[e^{-u\Delta_M}]\le\mathcal G_M(u)$ to obtain
\begin{equation}
d_M\ge -\frac{1}{u}\ln \mathcal{G}_M(u)=-\frac{1}{u}\Phi_M(u).
\end{equation}
Taking $\liminf_{M\to\infty}$ and invoking Eq.~\eqref{eq:scalinglimit} yields inequality~\eqref{eq:liminf_dM_u}.

\subsection{Proof of Eq.~\eqref{eq:trial_mean_defect}}
\label{app:proof_trial_mean_defect}

It is enough first to treat profiles whose Fourier amplitude is smooth and compactly supported inside the window, because these profiles are dense in the finite-second-moment domain.
Let $\phi\in C_c^\infty((-1,1))$ be real and even, $\norm{\phi}_2=1$, and set
\begin{equation}
g(\xi)=\frac{1}{\sqrt{2\pi}}\int_{-1}^{1}\phi(p)e^{\ii p\xi}\dd p.
\label{eq:unitary_inverse_app}
\end{equation}
With $p_j=(2j+1)/M$ and $h_M=2/M$, define
\begin{align}
\alpha_j^{(M)}&:=\sqrt{2h_M}\,\phi(p_j),\nonumber\\
\psi_{2j}^{(M,0)}&:=(-1)^j\alpha_j^{(M)},
&\psi_{2j+1}^{(M,0)}&:=0.
\label{eq:site_trial_unnormalized}
\end{align}
All remaining components are zero.
Since $\phi$ is compactly supported in $(-1,1)$, this state belongs to $\mathcal S_M$ for all large $M$.
Its norm $N_M:=\norm{\psi^{(M,0)}}$ satisfies
\begin{equation}
N_M^2=2h_M\sum_{j=0}^{N_{\rm c}-1}|\phi(p_j)|^2\longrightarrow1
\label{eq:NM_to_1}
\end{equation}
by the midpoint Riemann sum, and we set $\psi^{(M)}=\psi^{(M,0)}/N_M$.

On $D_M$ in \eqref{eq:WM_def}, define
\begin{equation}
g_M^{(0)}(\xi):=M^{-1/2}c_{\psi^{(M,0)}}\left(\frac{\pi}{2}+\frac{\xi}{M}\right).
\end{equation}
The cosine-sector identity gives the exact midpoint inverse transform
\begin{equation}
g_M^{(0)}(\xi)=\frac{h_M}{\sqrt{2\pi}}
\sum_{p\in\{\pm p_j\}}\phi(p)e^{\ii p\xi}.
\label{eq:gM0_discrete_inverse}
\end{equation}
Because $\phi$ is compactly supported in $(-1,1)$, for all sufficiently large $M$ the finite sum in Eq.~\eqref{eq:gM0_discrete_inverse} may be extended to the full midpoint lattice $p=(2j+1)/M$, $j\in\mathbb Z$, without changing its value.
Since $g$ is Schwartz, Poisson summation then yields, uniformly on $D_M$ and for every $N>0$,
\begin{align}
g_M^{(0)}(\xi)&=\sum_{\ell\in\mathbb Z}(-1)^\ell g(\xi-\pi M\ell),\nonumber\\
\sup_{\xi\in D_M}|g_M^{(0)}(\xi)-g(\xi)|&=O(M^{-N}).
\label{eq:poisson_midpoint}
\end{align}
Consequently,
\begin{equation}
\int_{D_M}\xi^2|g_M^{(0)}(\xi)-g(\xi)|^2\dd\xi\longrightarrow0.
\label{eq:weighted_gM_conv}
\end{equation}

Changing variables $k=\pi/2+\xi/M$ in the exact mean defect gives
\begin{align}
\E_{\psi^{(M)}}[\Delta_M]
&=\frac{1}{N_M^2}\int_{D_M}m_M(\xi)|g_M^{(0)}(\xi)|^2\dd\xi,\nonumber\\
m_M(\xi)&:=M^2\left(2-2\cos\frac{\xi}{M}\right).
\label{eq:exact_defect_gM}
\end{align}
Here $0\le m_M(\xi)\le\xi^2$ and $m_M(\xi)\to\xi^2$ pointwise.
Equation~\eqref{eq:weighted_gM_conv}, Cauchy-Schwarz, and dominated convergence therefore give
\begin{equation}
\lim_{M\to\infty}\E_{\psi^{(M)}}[\Delta_M]
=\int_{\mathbb R}\xi^2|g(\xi)|^2\dd\xi,
\end{equation}
which is Eq.~\eqref{eq:trial_mean_defect} for the trial profiles used in the main proof.
Real even functions in $C_c^\infty((-1,1))$ approximate the normalized Dirichlet ground state in $H_0^1((-1,1))$, so this family is sufficient for the matching upper bound.

\subsection{Proof of Eq.~\eqref{eq:varC1}}
\label{app:proof_varC1}

Recall the variational form \eqref{eq:var_lambda}.
Define the shorthand quantities
\begin{equation}
I_g(u)=\int_{\mathbb{R}} e^{-u\xi^2}\abs{g(\xi)}^2\dd \xi,\ \ 
\phi_g(u)=\ln I_g(u).
\end{equation}
Since $\lambda(u)=\max_{\norm{g}_2=1} I_g(u)$ and $\ln$ is increasing, we have the exact identity
\begin{equation}
\Phi(u)=\ln\lambda(u)=\max_{\norm{g}_2=1}\phi_g(u).
\label{eq:Phi_sup}
\end{equation}
If $g$ has finite second moment, then
\begin{equation}
\phi_g'(0^+)=-\int_{\mathbb{R}}\xi^2\abs{g(\xi)}^2\dd \xi.
\label{eq:phi_prime_0}
\end{equation}
At $u=0$, however, every normalized profile maximizes $I_g(0)=1$, so we do not differentiate the optimized value through this degenerate set.

Let $I=(-1,1)$, let $f=(2\pi)^{-1/2}\widehat g\in L^2(I)$, and denote its zero extension to $\mathbb R$ by $F$.
Plancherel gives
\begin{equation}
\langle f,\mathsf S_u f\rangle
=\frac{1}{2\pi}\int_{\mathbb R}e^{-ur^2}|\widehat F(r)|^2\dd r.
\end{equation}
Hence
\begin{equation}
\nu(u):=\frac{1-\lambda(u)}{u}
=\inf_{\norm{f}_{L^2(I)}=1}\frac{1}{2\pi}
\int_{\mathbb R}\frac{1-e^{-ur^2}}{u}|\widehat F(r)|^2\dd r.
\label{eq:nu_variational}
\end{equation}
For the normalized Dirichlet ground state $f_1(p)=\cos(\pi p/2)$, its zero extension belongs to $H^1(\mathbb R)$, and dominated convergence gives
\begin{equation}
\limsup_{u\downarrow0}\nu(u)\le\int_{-1}^{1}|f_1'(p)|^2\dd p=\frac{\pi^2}{4}.
\label{eq:nu_limsup}
\end{equation}

For the converse bound, choose a sequence $u_n\downarrow0$ that realizes $\liminf_{u\downarrow0}\nu(u)$.
For each $n$, choose a normalized $f_n$ in Eq.~\eqref{eq:nu_variational}, with zero extension $F_n$, such that
\begin{equation}
\frac{1}{2\pi}\int_{\mathbb R}\frac{1-e^{-u_nr^2}}{u_n}|\widehat F_n(r)|^2\dd r
\le \nu(u_n)+\frac{1}{n}.
\label{eq:near_minimizer_energy}
\end{equation}
The upper bound \eqref{eq:nu_limsup} shows that these energies remain bounded.
For fixed $R>0$ and all large $n$,
\begin{equation}
\frac{1}{2\pi}\int_{|r|>R}|\widehat F_n(r)|^2\dd r
\le \frac{C}{(1-e^{-1})R^2}.
\label{eq:Fourier_tail_bound}
\end{equation}
Indeed, $(1-e^{-u_nr^2})/u_n\ge(1-e^{-1})R^2$ when $|r|>R$ and $u_nR^2\le1$.
The functions $F_n$ have support in the fixed interval $[-1,1]$.
For a translation $\tau_hF_n(p)=F_n(p+h)$, the Plancherel theorem gives
\begin{equation*}
\norm{\tau_hF_n-F_n}_2^2
=\frac{1}{2\pi}\int_{\mathbb R}|e^{\ii rh}-1|^2|\widehat F_n(r)|^2\dd r.
\end{equation*}
After splitting the integral at $|r|=R$, the low-frequency part is bounded by $h^2R^2$, while Eq.~\eqref{eq:Fourier_tail_bound} controls the high-frequency part uniformly in $n$.
First taking $R$ large and then $h$ small proves uniform translation continuity.
The Riesz-Kolmogorov criterion~\cite{HancheOlsenHolden2010KolmogorovRiesz} therefore gives, after passing to a subsequence, $F_n\to F$ strongly in $L^2(\mathbb R)$, where $\norm{F}_2=1$ and $F=0$ outside $[-1,1]$.
For each fixed $R>0$, Eq.~\eqref{eq:near_minimizer_energy} gives
\begin{equation}
\nu(u_n)+\frac{1}{n}\ge\frac{1}{2\pi}\int_{|r|\le R}
\frac{1-e^{-u_nr^2}}{u_n}|\widehat F_n(r)|^2\dd r.
\end{equation}
On $|r|\le R$, the multiplier converges uniformly to $r^2$, while $\widehat F_n\to\widehat F$ strongly in $L^2(\mathbb R)$ by the Plancherel theorem.
It follows that
\begin{equation}
\liminf_{n\to\infty}\nu(u_n)
\ge\frac{1}{2\pi}\int_{|r|\le R}r^2|\widehat F(r)|^2\dd r.
\end{equation}
Letting $R\to\infty$ yields
\begin{equation}
\liminf_{n\to\infty}\nu(u_n)
\ge\frac{1}{2\pi}\int_{\mathbb R}r^2|\widehat F(r)|^2\dd r.
\end{equation}
Thus $F\in H^1(\mathbb R)$. Its continuous representative vanishes on $\mathbb R\setminus[-1,1]$, so $F(\pm1)=0$ and $F|_I\in H_0^1(I)$.
The Dirichlet Poincar\'e inequality gives
\begin{equation}
\frac{1}{2\pi}\int_{\mathbb R}r^2|\widehat F(r)|^2\dd r
=\int_{-1}^{1}|F'(p)|^2\dd p\ge\frac{\pi^2}{4}.
\end{equation}
Together with \eqref{eq:nu_limsup}, this proves
\begin{equation}
\lambda(u)=1-\frac{\pi^2}{4}u+o(u),
\qquad
\Phi(u)=-\frac{\pi^2}{4}u+o(u).
\end{equation}
The same Fourier representation shows that finite second moment is equivalent to $f\in H_0^1(I)$ and
\begin{equation}
\int_{\mathbb R}\xi^2|g(\xi)|^2\dd\xi=\int_{-1}^{1}|f'(p)|^2\dd p.
\end{equation}
Minimizing the right-hand side over normalized $f\in H_0^1(I)$ proves Eq.~\eqref{eq:varC1} and gives $C_1=\pi^2/4$.
\section{Numerical methods and convergence checks}
\label{app:numerics}
We summarize the numerical procedures used for the figures and list numerical convergence checks for the discretizations of $A^{(M)}$, $A_u^{(M)}$, and the continuum operator $T_u$.
We compute $A^{(M)}$ using the closed form \eqref{eq:Aclosed}.
The matrix is real symmetric, so we use a symmetric eigensolver and extract the largest eigenvalue and corresponding eigenvector.
The computed sequence $v_{\max}(M)$ is nondecreasing in $M$ and remains below $2$ in the numerical convention $J=1$, while the estimator $M^2(2-v_{\max}(M))$ approaches $\pi^2/4$.
For Fig.~\ref{fig:coherence}, the coherent curve is computed from the same diagonalization of $A^{(M)}$.
The site-diagonal reference value is not fitted; it is the closed-form value $16/(3\pi)$ in Eq.~\eqref{eq:incoherent_reference}.
For small $M$, the closed-form matrix agrees with direct quadrature of Eq.~\eqref{eq:Adef} to numerical precision.
We build $A_u^{(M)}$ from \eqref{eq:AuM} using a near-peak change of variables $k=\pi/2+\xi/M$ and truncate to $\xi\in[-\xi_{\max},\xi_{\max}]$.
On a uniform $\xi$ grid we approximate the integral by a Riemann sum, which converges rapidly because the integrand is smooth and exponentially localized for fixed $u>0$.
In the figures we use $\xi_{\max}=12$ with $N_\xi=3001$-$4001$ points depending on the plot.
We verified convergence by increasing $\xi_{\max}$ and doubling $N_\xi$ and checking stability of the largest eigenvalue to a prescribed tolerance.

Several finite-support matrices in the main text can be assembled from the sine-sine integral form
\begin{equation}
(K_f^{(M)})_{mn}=\frac{2}{\pi}\int_0^\pi \sin[(m+1)k]\sin[(n+1)k] f(k) \dd k,
\label{eq:gen_matrix_gk}
\end{equation}
with a smooth weight $f(k)$.
Using the product-to-sum identity gives the Toeplitz-Hankel decomposition
\begin{equation}
(K_f^{(M)})_{mn}=t_{|m-n|}-h_{m+n},
\label{eq:TH_num}
\end{equation}
where
\begin{align}
 &t_r =\frac{1}{\pi}\int_0^\pi \cos(rk) f(k) \dd k,\\
 &h_q =\frac{1}{\pi}\int_0^\pi \cos[(q+2)k] f(k) \dd k.
\end{align}
This reduces matrix assembly to one-dimensional quadrature for the coefficient arrays $\{t_r\}_{r=0}^{M-1}$ and $\{h_q\}_{q=0}^{2M-2}$.
For example, $f(k)=e^{2Js\sin k}$ gives $B^{(M)}(s)$ in Sec.~\ref{sec:finiteM}, and the weight in \eqref{eq:AuM} gives $A_u^{(M)}$.

We turn to the numerical evaluation of the continuum scaling function $\lambda(u)$.
This requires discretizing the limiting integral operator representation introduced in Sec.~\ref{sec:continuum}, and we use a standard quadrature-based scheme for that purpose.
We compute $\lambda(u)$ using the Fourier-side operator $\mathsf{S}_u$ with kernel \eqref{eq:Su_kernel} on $[-1,1]$.
We discretize it by a Nystr\"om method~\cite{Nystrom1930IntegralEquations} with Gauss-Legendre nodes $p_i$ and weights $\rho_i$, forming the symmetric matrix
\begin{equation}
(\mathsf{S}_u)_{ij}\approx \sqrt{\rho_i}\mathcal{K}_u(p_i,p_j)\sqrt{\rho_j}.
\end{equation}
The largest eigenvalue converges rapidly because the kernel is smooth for $u>0$.
The computed values are monotone in $u$ and obey the bound $\lambda(u)\le 1$.

\section{Parity decomposition and pairwise plateaus}
\label{app:parity_plateau}
We record two finite-$M$ consequences of the site-parity structure.  The first gives the exact pairwise plateau in the mean-drift optimum.  The second explains why the biased optimizer in Fig.~\ref{fig:optbias} has the same support parity.

\subsection{Pairwise plateau for the mean-drift matrix}

The closed form \eqref{eq:Aclosed} implies that $A^{(M)}_{mn}=0$ whenever $m+n$ is odd, so $A^{(M)}$ decomposes into even-site and odd-site blocks.  For $r\ge1$, define the two $r\times r$ blocks
\begin{equation}
(A_{\rm e}^{(r)})_{ij}=A_{2i,2j}^{(2r)},\quad
(A_{\rm o}^{(r)})_{ij}=A_{2i+1,2j+1}^{(2r)},
\label{eq:CS_blocks}
\end{equation}
where $i,j=0,\ldots,r-1$.
With the convention that $A_{\rm o}^{(0)}$ is absent, the decomposition reads
\begin{equation}
A^{(2r-1)}\cong A_{\rm e}^{(r)}\oplus A_{\rm o}^{(r-1)},\quad
A^{(2r)}\cong A_{\rm e}^{(r)}\oplus A_{\rm o}^{(r)},
\label{eq:pair_block_decomp}
\end{equation}
where $\cong$ denotes equality up to a basis permutation.

It remains to compare the two blocks of the same size.  Using \eqref{eq:Aclosed}, one obtains
\begin{align}
(A_{\rm e}^{(r)}-A_{\rm o}^{(r)})_{ij}
&=\frac{16J}{\pi(2i+2j+1)(2i+2j+3)(2i+2j+5)}
\notag\\
&=\frac{2J}{\pi}
\int_0^1 x^{2i+2j}(1-x^2)^2\dd x .
\label{eq:CS_difference}
\end{align}
Therefore, for any nonzero $a=(a_0,\ldots,a_{r-1})\in\mathbb C^r$,
\begin{equation}
a^*(A_{\rm e}^{(r)}-A_{\rm o}^{(r)})a
=\frac{2J}{\pi}
\int_0^1\left|\sum_{i=0}^{r-1}a_i x^{2i}\right|^2(1-x^2)^2\dd x>0.
\label{eq:CS_positive}
\end{equation}
Thus $A_{\rm e}^{(r)}-A_{\rm o}^{(r)}$ is positive definite and $\lambda_{\max}(A_{\rm e}^{(r)})>\lambda_{\max}(A_{\rm o}^{(r)})$.  Since $A_{\rm o}^{(r-1)}$ is a principal submatrix of $A_{\rm o}^{(r)}$, $\lambda_{\max}(A_{\rm o}^{(r-1)})\le\lambda_{\max}(A_{\rm o}^{(r)})$.  Equation~\eqref{eq:pair_block_decomp} then gives
\begin{equation}
\lambda_{\max}(A^{(2r-1)})=\lambda_{\max}(A^{(2r)})=\lambda_{\max}(A_{\rm e}^{(r)}),
\label{eq:pair_plateau_proof}
\end{equation}
which proves \eqref{eq:pairwise_plateau_main}.

\subsection{Parity selection for weighted matrices}

We next use the same sine-sine matrix class as in \eqref{eq:gen_matrix_gk}.  Let $w:[0,1]\to\mathbb R_+$ be nonincreasing, set $f_w(k)=w(\cos^2 k)$, and write $K_w^{(M)}:=K_{f_w}^{(M)}$.  Equivalently,
\begin{equation}
(K_w^{(M)})_{mn}=\frac{2}{\pi}\int_0^\pi
\sin[(m+1)k]\sin[(n+1)k]w(\cos^2 k)\dd k .
\label{eq:Kw_def}
\end{equation}
This class includes the mean-drift matrix with $w(y)=2J\sqrt{1-y}$ and the biased matrix $A_u^{(M)}$ with
\begin{equation}
w_{u,M}(y)=\exp[-uM^2(2-2\sqrt{1-y})].
\label{eq:w_uM}
\end{equation}
Since $w(\cos^2 k)$ is invariant under $k\mapsto\pi-k$, $K_w^{(M)}$ decomposes into even-site and odd-site blocks.

We compare blocks of the same size.  Set $x=\cos k$ and $y=x^2$.  Using $\sin[(n+1)k]=\sin k\,U_n(x)$, where $U_n$ is the Chebyshev polynomial of the second kind, an even-site vector corresponds to a polynomial $Q(y)$ of degree at most $r-1$, while an odd-site vector of the same dimension corresponds to $xQ(y)$.  These parametrizations are invertible because $U_{2j}(x)$ and $U_{2j+1}(x)/x$ are polynomials of degree $j$ in $y$.
With
\begin{equation}
\dd\nu(y)=\frac{2}{\pi}\frac{\sqrt{1-y}}{\sqrt y}\dd y,
\end{equation}
the two Rayleigh quotients associated with the same polynomial $Q$ are
\begin{align}
R_{\rm e}[Q]&=\frac{\int_0^1w(y)|Q(y)|^2\dd\nu(y)}{\int_0^1|Q(y)|^2\dd\nu(y)},\notag\\
R_{\rm o}[Q]&=\frac{\int_0^1yw(y)|Q(y)|^2\dd\nu(y)}{\int_0^1y|Q(y)|^2\dd\nu(y)}.
\label{eq:evenodd_RQ}
\end{align}
If $\mu_Q$ is the probability measure proportional to $|Q(y)|^2\dd\nu(y)$, then $\mathbb E_{\mu_Q}[y]>0$ and
\begin{align}
R_{\rm o}[Q]-R_{\rm e}[Q]
&=\frac{\operatorname{Cov}_{\mu_Q}(y,w(y))}{\mathbb E_{\mu_Q}[y]}\le0,
\label{eq:covariance_order}
\end{align}
because
\begin{align}
&\operatorname{Cov}_{\mu_Q}(y,w(y))\nonumber\\
&=\frac12\int_0^1\int_0^1(y-z)[w(y)-w(z)]\dd\mu_Q(y)\dd\mu_Q(z)\le0 .
\end{align}
The inequality is strict when $w$ is strictly decreasing and $Q$ is nonzero.  For $M=2r$ this comparison applies directly.  For $M=2r-1$, the odd-site block has dimension $r-1$ and is a principal submatrix of the odd block of size $r$ for the same weight.  It follows that, for every fixed $u>0$, the principal eigenvalue of $A_u^{(M)}$ lies in the even-site block.  Thus the $u=1$, $M=200$ optimizer in Fig.~\ref{fig:optbias} is parity selected for the same reason as the mean-drift optimizer.

The pairwise identity \eqref{eq:pairwise_plateau_main}, however, is a stronger finite-size statement special to the mean-drift matrix $A^{(M)}$.  It does not extend to $A_u^{(M)}$, because the weight \eqref{eq:w_uM} itself depends on $M$.  This exact plateau is a finite-size parity effect and does not alter the large-$M$ asymptotic law \eqref{eq:vmax_final}.

\bibliography{ref}

\end{document}